# AgriTrust: a Federated Semantic Governance Framework for Trusted Agricultural Data Sharing

Ivan Bergier
**Embrapa Digital Agriculture, Campinas, SP, Brazil.**
ivan.bergier@embrapa.br

**Abstract**

The potential of agricultural data (AgData) to drive efficiency and sustainability is stifled by the "AgData Paradox": a pervasive lack of trust and interoperability that locks data in silos, despite its recognized value. This paper introduces AgriTrust, a federated semantic governance framework designed to resolve this paradox. AgriTrust integrates a multi-stakeholder governance model, built on pillars of Data Sovereignty, Transparent Data Contracts, Equitable Value Sharing, and Regulatory Compliance, with a semantic digital layer. This layer is realized through the AgriTrust Core Ontology, a formal OWL ontology that provides a shared vocabulary for tokenization, traceability, and certification, enabling true semantic interoperability across independent platforms. A key innovation is a blockchain-agnostic, multi-provider architecture that prevents vendor lock-in. The framework's viability is demonstrated through case studies across three critical Brazilian supply chains: coffee (for EUDR compliance), soy (for mass balance), and beef (for animal tracking). The results show that AgriTrust successfully enables verifiable provenance, automates compliance, and creates new revenue streams for data producers, thereby transforming data sharing from a trust-based dilemma into a governed, automated operation. This work provides a foundational blueprint for a more transparent, efficient, and equitable agricultural data economy.

## 1. Introduction

Brazilian agribusiness is a cornerstone of the global food system and a frontier of technological adoption. The sector's digital transformation is generating vast volumes of agricultural data (AgData), which holds immense potential to drive unprecedented gains in efficiency, sustainability, and value creation [1,2]. This potential has been rendered increasingly urgent by regulations like the EU Deforestation-Free Regulation (EUDR), which make verifiable AgData a concrete prerequisite for market access.

However, this opportunity is stifled by a pervasive "AgData Paradox" [3-6]. Formally, the paradox reflects the observed contradiction in the agricultural data economy wherein there is a universal recognition of the high potential value of data sharing among stakeholders, yet this potential remains systematically unrealized due to a pervasive triad of barriers. Firstly, a trust deficit, stemming from producers' fears over data sovereignty and unfair value appropriation. Secondly, a lack of technical interoperability, resulting from heterogeneous data sources and formats. Finally, an absence of standardized, fair governance, leading to unclear rules for data ownership, usage consent, and benefit sharing. This paradox manifests as a landscape of fragmented data silos, which directly impedes critical outcomes like verifiable supply chain traceability and automated regulatory compliance.

Current technological solutions often address only part of the problem. Isolated traceability platforms or singular blockchain implementations can ensure data integrity but frequently fail to resolve the foundational governance challenges: Who owns the data? Under what conditions can it be used? How are the benefits shared? [5-8]. Without a trusted, standardized, and fair governance framework, these solutions struggle to achieve widespread adoption and, crucially, interoperability across the ecosystem, risking the creation of new, more technologically advanced silos.

This limitation highlights a significant research gap at the intersection of data governance, semantic technologies, and distributed systems. While substantial work exists in each field independently, from the FAIR and CARE principles to OWL and blockchain [9-12], their integration into a cohesive, operational framework for agriculture remains nascent. The critical missing piece is a foundational layer that seamlessly intertwines enforceable governance rules with machine-interpretable semantics.

To bridge this gap, this paper proposes AgriTrust, a federated semantic governance framework for trusted agricultural data sharing. Our primary contributions are:

i) An Integrated Governance-Digital Framework that combines a multi-stakeholder governance model with a semantic digital layer based on a formal OWL ontology;
ii) The AgriTrust Core Ontology, a comprehensive OWL ontology providing a shared vocabulary for agricultural tokenization, traceability, and certification;
iii) A Blockchain-Agnostic, Multi-Provider Architecture that prevents vendor lock-in and fosters a competitive ecosystem; and
iv) Proof-of-Concept Validation via Multi-Commodity Case Studies: a demonstration of the framework's viability through detailed case studies across three critical Brazilian supply chains, supported by a complete ontology specification (Appendix A), practical data contracts and queries (Appendix B), and platform integration codes (Appendix C).

The remainder of this paper is organized as follows. Section 2 reviews the relevant background and related work. Section 3 details the AgriTrust framework, encompassing its governance model, core ontology, and technical architecture. Section 4 presents the implementation and validation through multi-commodity case studies. Section 5 discusses the results, framework limitations, and a comparative analysis. The final section highlights main conclusions and outlines directions for future work.

**2. Background and Related Work**

The challenge of enabling trusted data sharing in agriculture sits at the intersection of several fields of study. This section reviews the relevant literature and existing initiatives that form the foundation for the proposed AgriTrust framework, while also delineating its unique contributions.

*2.1 Data Governance and Sovereignty in Agriculture*

The principles of data governance provide the scaffolding for managing data assets effectively. General frameworks, such as those promoted by the Global Data Alliance, emphasize principles like transparency, accountability, and fairness [8,9]. However, the agricultural context demands models that are more specific. The concept of data sovereignty, the idea that data is subject to the laws and governance structures of the nation where it is collected, takes on a critical dimension at the farm level, where producers seek to maintain control over data generated on their land [5,13].

This is articulated by the CARE Principles for Indigenous Data Governance (Collective Benefit, Authority to Control, Responsibility, and Ethics) [10], which have emerged as a crucial complement to the well-known FAIR principles (Findable, Accessible, Interoperable, Reusable) [9]. While FAIR focuses on data openness and reusability by machines, CARE shifts the focus to people-centric governance, emphasizing the rights of data originators. AgriTrust aligns with this view, embedding producer sovereignty as a non-negotiable pillar, thereby addressing a primary barrier to participation identified in studies on private AgData ecosystems [13].

*2.2 Semantic Technologies and Interoperability*

Overcoming syntactic and semantic heterogeneity is a prerequisite for meaningful data sharing. Ontologies, formalized using standards like the OWL, provide a shared vocabulary and a machine-interpretable understanding of a domain's concepts and their relationships [11]. In agribusiness, significant efforts have been made to create common semantic frameworks [14-16].

Initiatives such as the Agricultural Data Exchange (ADE) and the work of ISO/TC 347 on data-driven agri-food systems aim to create standardized models for supply chain data [14]. Repositories like AgroPortal host numerous ontologies and vocabularies for the agri-food sector, demonstrating the community's recognition of the interoperability challenge [15]. A recent literature review comprehensively addresses the use of semantic resources and explainable AI to overcome the challenges of data integration, interoperability, and explainability in smart agricultural systems [16]. AgriTrust builds upon this foundation but introduces a dedicated OWL ontology focused explicitly on the concepts of tokenization and governed data sharing, providing a semantic layer that allows different platforms to interpret a concept like a "Token representing a sustainable coffee batch" in an identical manner. This enables true interoperability beyond simple data format conversion.

*2.3 Tokenization, Blockchain, and Digital Assets*

Tokenization, the process of creating a cryptographic, digital representation of a physical or intangible asset on a digital ledger, is a mechanism for traceability and certification. A substantial body of research has explored blockchain-based traceability for various supply chains, from agri-food [17] to pharmaceuticals [18]. However, many implementations are siloed [4], relying on proprietary data models and single-chain architectures [8] that limit broader ecosystem integration [16].

These systems often prioritize immutability and transparency but fail to adequately address the granular, legally-enforceable access control required for sensitive commercial and operational data. AgriTrust formalizes the key concepts of a Tokenization-as-a-Service (TaaS) platform, explicitly linking the technical process of tokenization to a governance model that defines the rules for who can create tokens, under what conditions, and what rights they confer. This enhances the trustworthiness of the digital asset itself by grounding it in a legal and operational framework, a layer often missing from purely technical blockchain solutions.

*2.4 Data Spaces and Federated Architectures*

The concept of data spaces, exemplified by architectures such as the International Data Spaces (IDS) [19], provides a robust blueprint for creating sovereign, federated data-sharing environments. A data space is not a centralized database but a governed infrastructure that connects distributed data sources through common rules and standards. Participants retain control over their data, sharing it under predefined terms and conditions, a philosophy that deeply resonates with the AgriTrust vision.

The technological architecture of AgriTrust, with its standardized application programming interfaces (APIs), core ontology, and federated query capability, can be viewed as an instantiation of a specialized agricultural data space for traceability and certification. However, while IDS provides a general-purpose reference model, AgriTrust contributes a domain-specific implementation for Brazilian agribusiness, complete with a ready-to-use ontology and a detailed governance model tailored to the sector's unique actors, processes, and regulatory requirements.

*2.5 Synthesis and Identified Gap*

The related work reveals a fragmented landscape. Significant progress has been made in individual domains: semantic ontologies exist for specific commodities, blockchain platforms offer robust traceability, and data space architectures provide a vision for federation. However, a critical gap remains in the integration of these components into a cohesive, operational whole.

Firstly, existing ontologies often lack the conceptual machinery to model governance artifacts like *Data Contracts*. Conversely, governance frameworks frequently remain abstract, without a concrete technical implementation. Secondly, many blockchain solutions create new walls around data, focusing on internal consistency rather than cross-platform interoperability. Finally, many initiatives prioritize data accessibility for consumers and certifiers without providing an equally robust mechanism for ensuring data sovereignty and fair value sharing for producers.

AgriTrust addresses this gap by proposing a unified framework that does not merely combine but deeply integrates a multi-stakeholder governance model with a semantic digital layer within a blockchain-agnostic, federated architecture. It moves beyond conceptual models and siloed prototypes to offer a practical, scalable blueprint for transforming the AgData Paradox into a solvable coordination problem.

**3. The AgriTrust Framework: Integrated Governance and Semantic Architecture**

This section presents the AgriTrust framework, an integrated model designed to resolve the AgData Paradox by intertwining a robust multi-stakeholder governance framework with a semantic digital layer. The core innovation lies in ensuring that technical interoperability is not an afterthought but is guided and enforced by clear, fair, and transparent rules of engagement from the outset.

*3.1 Framework Overview and Design Principles*

The AgriTrust framework is architected as a federated ecosystem, not a centralized platform. In this context, a federated model refers to an architectural paradigm that connects a distributed network of independent, autonomous nodes (e.g., TaaS Platforms, data sources) through a common set of governance rules and technical standards. Unlike a centralized system that consolidates data and control into a single entity, a federated system: preserves sovereignty (each participant retains control over their own data and infrastructure, ensures interoperability (a shared semantic layer or the Core Ontology and standardized interfaces enable seamless communication and data exchange between nodes), and prevents monopoly (the ecosystem is designed for multi-provider participation, avoiding vendor lock-in and fostering innovation through competition).

This model creates a cohesive data-sharing environment without requiring data to be physically centralized, thereby directly upholding the principle of data sovereignty. It connects distributed data sources, managed by various *Platform Providers*, through a common set of governance rules and semantic standards. The high-level architecture is depicted in Figure 1.

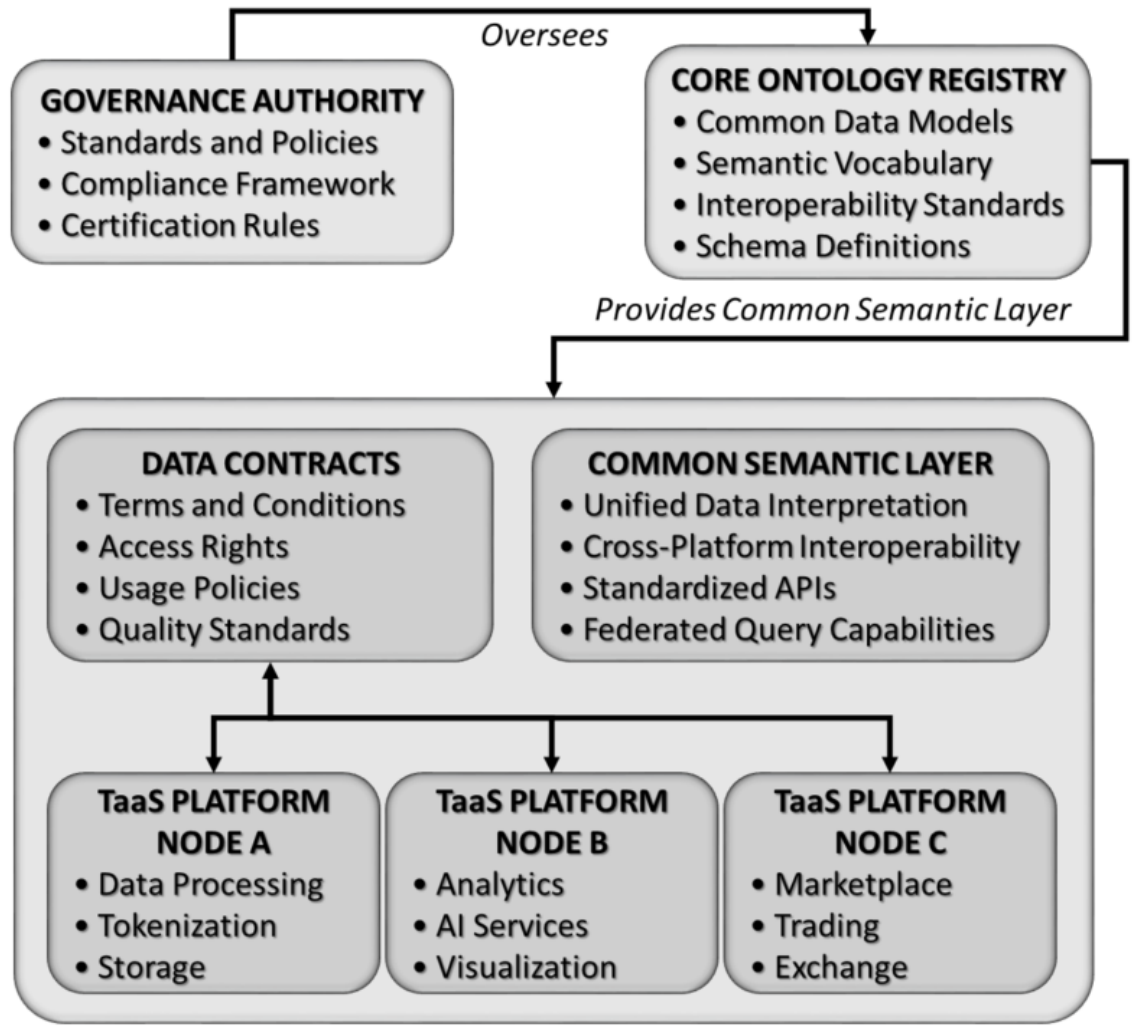

Figure 1. Overview of the integrated governance and digital AgData sharing ecosystem, showing Governance Authority, Core Ontology Registry, and multiple TaaS Platform Nodes (A, B, C) interacting via Data Contracts in a Common Semantic Layer.

The design is guided by four core pillars or principles:

1. *Data Sovereignty* as the main foundation: producers and data originators retain ultimate ownership and control. The system is designed to enforce their authority over how data is classified, used, and shared.
2. *Governance-Led* technical implementation: The governance model defines the "rules of the game"; the digital infrastructure is built to execute these rules faithfully, making compliance the default.
3. *Federated*, not centralized: The ecosystem thrives on the participation of multiple, independent Platform Providers and Blockchain Providers, avoiding monopolistic control and fostering innovation through competition and specialization.
4. *Semantic Interoperability* as a prerequisite: Trusted data sharing requires a shared, unambiguous understanding of meaning. A formal ontology provides this common language, enabling both humans and machines to interpret data consistently across platform boundaries.

*3.2 Governance Model*

The governance model establishes the rules, roles, and processes that create a fair, transparent, and enforceable environment for all participants.

*3.2.1 Multi-Stakeholder Governance Authority*

A cornerstone of AgriTrust is the Governance Authority, an entity (typically a consortium comprising producer associations, industry representatives, certifiers, and government bodies) responsible for the ecosystem's integrity. Its mandates include:

- Maintaining and evolving the core governance framework;
- Managing the version-controlled Core Ontology Registry;
- Accrediting Platform and Blockchain Providers; and
- Providing a dispute resolution mechanism.

*3.2.2 Core Governance Pillars*

The framework's stability is built upon four pillars, instantiated through legal and technological instruments:

1. Producers and data originators retain ultimate authority, technically enforced via access control mechanisms in Data Contracts.
2. Machine-readable Data Contracts (detailed in Section 3.4.1) govern all data sharing transactions. These are digitally signed agreements that specify the purpose, duration, scope, and terms of data use, creating an auditable trail of all data-related transactions.
3. The model explicitly recognizes that data has economic value and promotes mechanisms for the fair distribution of benefits derived from shared data among all contributors. This can be operationalized through automated premium payments or access-to-service agreements encoded within Data Contracts.
4. The framework is designed to ensure that data sharing practices inherently adhere to relevant laws and norms, such as Brazil's *Lei Geral de Proteção de Dados* (LGPD) and international standards like the EU Deforestation-Free Regulation (EUDR).

*3.2.3 Roles and Responsibilities*

Clear roles delineate responsibilities and expectations within the ecosystem:

- Data Producer: The entity that originates data (e.g., a farmer or a cooperative providing yield, sensor, or management data). They are the sovereign owners of their data.

- Data Consumer: The entity that uses data under specific terms defined in a Data Contract (e.g., a certifier validating sustainable practices, a market requesting proof of origin).
- Platform Provider: Offers the technical infrastructure (e.g., a TaaS platform) for creating and managing digital tokens and enforcing Data Contracts. They must maintain neutrality and adhere to the governance rules.
- Certifier: A trusted third-party entity responsible for auditing and issuing compliance certificates (e.g., sustainable, deforestation-free) based on verifiable data accessed via Data Contracts.

### 3.3 Semantic Layer: The AgriTrust Core Ontology

The governance rules are operationalized by the digital framework, whose core is a shared OWL ontology. This ontology provides the common language that allows independent platforms to interoperate seamlessly, overcoming both syntactic and semantic heterogeneity.

### 3.3.1 Ontology Design Methodology

Developers and researchers are directed to Appendix A for the definitive, ready-to-use semantic foundation. It is presented in Turtle format and includes SHACL constraints to ensure data integrity and validation for implementing platforms. While the core concepts are summarized below and in Figures 2 and 3, the full ontology in Appendix A serves as the definitive, ready-to-use semantic foundation for developers seeking to build interoperable TaaS platforms. The AgriTrust Core Ontology reuses and aligns with well-established standards to ensure broad interoperability:

- PROV-O for modeling provenance, ensuring a clear chain of custody for all assets.
- SOSA/SSN for representing observations and measurements from sensors and manual inputs.
- GeoSPARQL for representing spatial geometries of farms and plots.
- AGROVOC and GS1 for aligning with domain-specific terminologies.

### 3.3.2 Core Classes and Properties

The ontology defines a standardized vocabulary for key concepts in agricultural tokenization and traceability. Its core classes and their primary relationships are summarized in Table 1, providing a foundational overview of the semantic model.

Table 1: Core Classes and Properties of the AgriTrust Ontology

| Core Class | Description | Key Properties |
|---|---|---|
|  |  |  |
| agt:Asset | A physical or digital asset that can be tokenized (e.g., a cow, a grain batch). | agt:ownedBy, agt:hasProvenance, agt:hasCertificate |
| agt:Token | A digital representation of an Asset on a ledger. | agt:represents, agt:registeredOnBlockchain, agt:governedBy |
| agt:Process | An activity that transforms/handles Assets (e.g., harvesting, transport). | agt:hasInput, agt:hasOutput |
| agt:Observation | An act of sensing or measuring (e.g., soil analysis, weight). | agt:observationValue, agt:observationDate |
| agt:DataContract | A machine-readable agreement governing data sharing. | agt:purpose, agt:validUntil, agt:coversAsset |
| agt:Certificate | A verifiable credential attesting to compliance or quality. | agt:certifiedBy, agt:standard |
| agt:Agent | A participant in the ecosystem (e.g., Producer, Consumer, Certifier). | (Role-specific properties) |

The relationships between these core classes form a graph model for representing supply chain knowledge. This structure is depicted conceptually in Figure 2, which shows how a tokenized asset is linked to its provenance, supporting evidence, and the governance contracts that control access to its data. This graph-based approach is what enables the complex, federated querying demonstrated later in the paper. These classes are interlinked through the object properties, creating a rich, graph-based model of the supply chain. For instance, an agt:Token agt:represents an agt:Asset, which agt:hasProvenance from a agt:Process and agt:hasCertificate issued by an agt:Certifier (Figure 2).

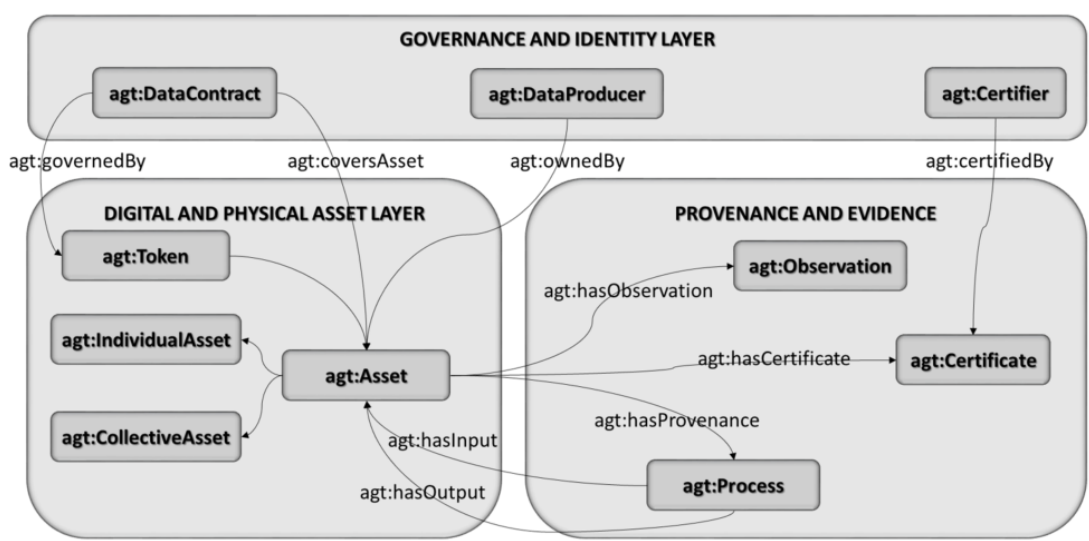

Figure 2. Core Relationships of the AgriTrust Ontology. This conceptual graph illustrates the key classes and their primary relationships, showing how a digital agt:Token represents a physical agt:Asset, whose provenance is tracked via agt:Process and evidenced by agt:Observation. The entire system is governed by machine-readable agt:DataContract.

*3.3.3 Multi-Provider Extensibility Pattern*

A critical feature of the AgriTrust Ontology is its design for extensibility. The core is kept minimal and stable. Platform Providers can define their own subclasses and properties for domain-specific concepts without breaking interoperability. For example, a beef platform can create a subclass beef:FinishingOperation of agt:Process, while a soy platform can create soy:YieldPerHectare as a subclass of agt:EfficiencyMetric. As long as all platforms commit to the core ontology, they retain the ability to query and understand the fundamental relationships across the ecosystem.

*3.4 Technical Architecture*

The integration of governance and semantics is realized through a federated technical architecture that embeds the rules directly into the data-sharing infrastructure.

*3.4.1 Core Components*

The architecture, illustrated in detail in Figure 3, is composed of the following key components:

- Core Ontology Registry: A central, version-controlled repository for the shared OWL ontology, maintained by the Governance Authority. All Platform Providers must synchronize with this registry to ensure a consistent semantic understanding across the ecosystem.
- TaaS Platforms: Independent platforms operated by cooperatives, agroindustries, or third-party providers. They implement the core ontology and are responsible for:
    - Tokenizing Assets (agt:Asset -> agt:Token).
    - Hosting and serving verifiable data about Assets and Processes.
    - Enforcing Data Contracts via smart contracts or secure, policy-enforced APIs.
    - Identity and Access Management (IAM): A decentralized system (e.g., based on Decentralized Identifiers - DIDs) for authenticating all participants (agents, platforms) within the ecosystem.
    - Query Interfaces: Standardized APIs (primarily SPARQL endpoints) that allow participants to perform federated queries across different TaaS Platforms using the common ontological terms. A portfolio of practical, validated SPARQL queries demonstrating complex use cases is provided in Appendix B.2.

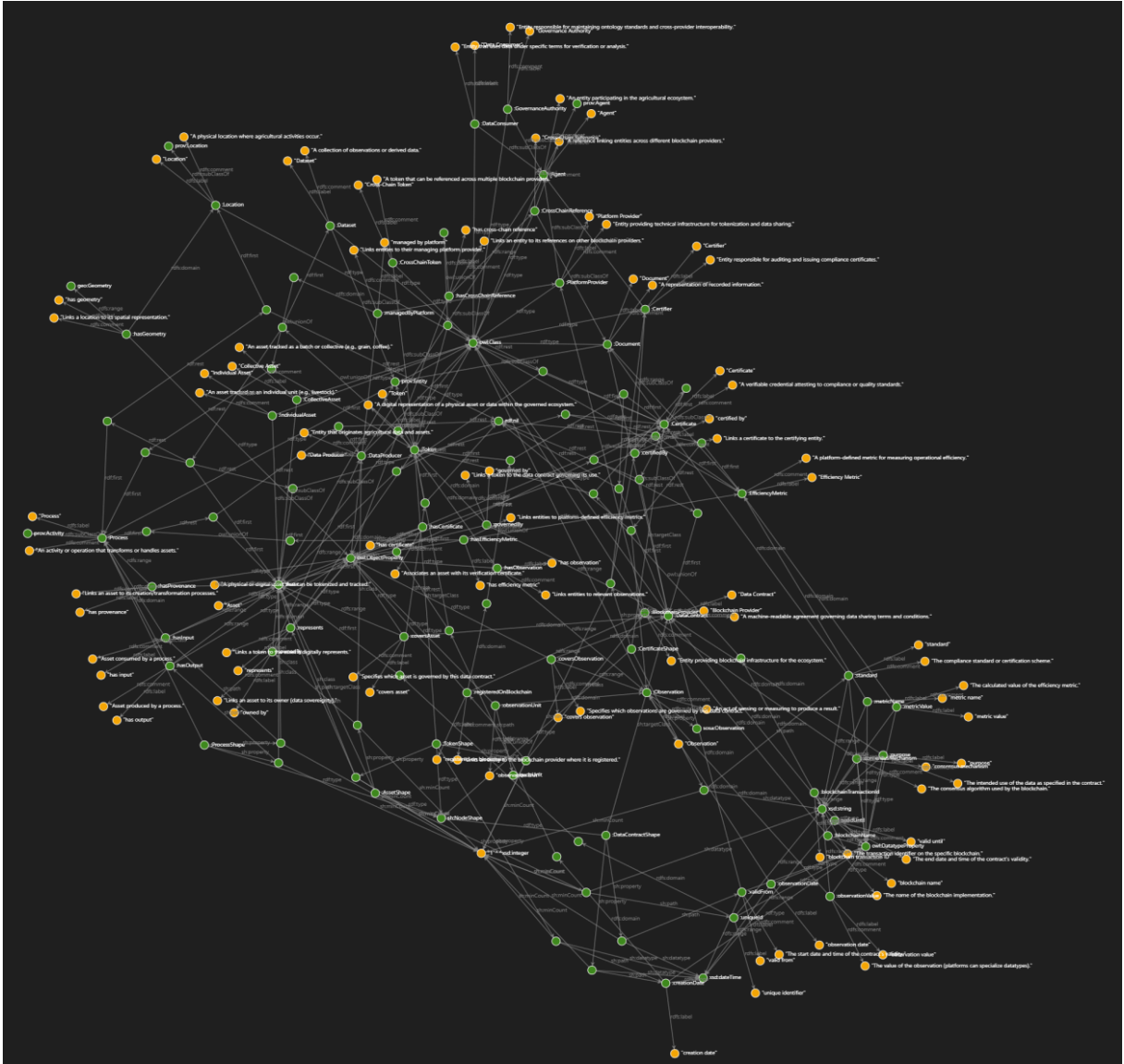

Figure 3. AgriTrust Federated Technical Architecture. This diagram illustrates the key components of the AgriTrust architecture and their interactions. The Governance Authority maintains the Core Ontology Registry. Accredited TaaS Platforms synchronize with this registry and provide services like tokenization and data hosting. Participants are authenticated via a decentralized IAM system. Data Consumers (e.g., Certifiers, Retailers) interact with the ecosystem through standardized Query Interfaces to execute federated queries across the TaaS Platforms, with all data access governed by machine-readable Data Contracts.

*3.4.2 Data Contracts: The Operational Heart*

The agt:DataContract is the pivotal entity that connects the governance and technical layers. It is a machine-readable manifestation of a legal agreement. A typical Data Contract instance specifies:

```
:DataContract_Certification a agt:DataContract ;
    dct:title "Sustainable Certification Data Agreement" ;
    agt:validFrom "2024-06-25T00:00:00Z"^^xsd:dateTime ;
```

```
    agt:validUntil "2024-12-25T00:00:00Z"^^xsd:dateTime ;
    agt:purpose "sustainable_certification_verification" ;
    agt:dataConsumer :Certifier_EuroSustainable ;
    agt:coversAsset :CoffeeBatch_2024_A ;
    agt:allowedUsage "compliance_audit" ;
    agt:prohibitedUsage "commercial_resale" .
```

Technologically, these contracts can be implemented as smart contracts on a blockchain or as policy documents in a secure API gateway. Their execution is automated: access is granted only when contract conditions are met and revoked upon expiration or breach.

For a more realistic scenario, consider a Data Contract for sustainable certification, which governs the sharing of specific farm data with a certification body, including clear usage restrictions and value-sharing terms:

```
:DataContract_SustainableCert_001 a agt:DataContract ;
  dct:title "Sustainable Certification Data Sharing Agreement" ;
  agt:validFrom "2024-06-25T10:00:00Z"^^xsd:dateTime ;
  agt:validUntil "2024-12-25T10:00:00Z"^^xsd:dateTime ;

  # Contract Parties
  prov:wasAttributedTo :CoffeeFarm_Santos ;       # Data Producer
  prov:wasGeneratedBy :Certifier_EuroSustainable ; # Data Consumer

  # Purpose and Scope
  agt:purpose "sustainable_certification_verification" ;
  agt:coversAsset :CoffeeBatch_2024_A ;
  agt:coversObservation :SoilAnalysis_2024, :WaterUsage_2024 ;

  # Usage Restrictions (Machine-Readable)
  agt:allowedUsage "compliance_verification_against_eu_sustainable_standard" ;
  agt:prohibitedUsage "commercial_resale, ai_training, marketing_activities" ;

  # Equitable Value Sharing
  agt:compensationType "premium_price_access" ;
  agt:compensationValue "0.15" ; # USD per kg premium

  # Technical Implementation
  agt:apiEndpoint "https://api.coop-platform.com/coffee/batch/2024A" ;
  agt:queryInterface "SPARQL" .
```

This contract exemplifies how the four governance pillars or principles (Section 3.1) are translated into executable code.

For a comprehensive view of real-world contract templates, including those for supply chain transparency with automated value-sharing, see Appendix B.1.

*3.4.3 Blockchain-Agnostic Implementation*

A key innovation of AgriTrust is its deliberate blockchain-agnosticism. The framework does not prescribe a single blockchain provider. Instead, the ontology includes properties to register entities on various ledgers:

```
:Token_Coffee_001 a agt:Token ;
    agt:represents :CoffeeBatch_2024_A ;
    agt:registeredOnBlockchain :BrazilAgriChain ;
    agt:blockchainTransactionId "tx_brazil_abc123" .

:Certificate_EUDR_001 a agt:Certificate ;
    agt:registeredOnBlockchain :EUCertChain ;
    agt:blockchainTransactionId "tx_eu_xyz789" .
```

This allows a token representing a Brazilian coffee batch to be registered on a national, low-cost blockchain (:BrazilAgriChain), while its EUDR compliance certificate is registered on a blockchain favored by European markets (:EUCertChain). The agt:hasCrossChainReference property can then link these entities, creating a verifiable, cross-chain provenance record without relying on a single, monolithic ledger.

It is important to clarify the scope of this blockchain-agnosticism. The AgriTrust framework is designed to be agnostic at the level of asset registration and provenance anchoring. The ontology provides the semantic layer to reference and link entities across different ledgers, as shown in the example. This approach successfully prevents vendor lock-in and allows participants to choose ledgers based on cost, jurisdiction, or performance for registering their tokens and certificates. However, the framework in its current form does not solve the more complex challenge of cross-chain logic execution, where a smart contract on one blockchain autonomously triggers a state change or payment on another. Such advanced interoperability, while a valuable area for future exploration, is outside the defined scope of the current semantic and governance model. The primary goal is to ensure semantic and referential interoperability across chains for traceability and certification, not to facilitate complex, cross-chain decentralized applications.

*3.4.4 Federated Query and Interoperability*

The true power of the semantic layer is realized through federated querying. For instance, a European retailer can assemble a due diligence report for EUDR compliance by querying across multiple platforms in a single, unified query:

```
PREFIX agt: <http://example.org/ns/agritrustcore#>
PREFIX geo: <http://www.opengis.net/ont/geosparql#>
PREFIX xsd: <http://www.w3.org/2001/XMLSchema#>
PREFIX prov: <http://www.w3.org/ns/prov#>

SELECT ?product ?farm ?geometry ?satelliteImage ?imageDate ?complianceStatus
WHERE {
    # Identify the product batch from a trader's platform
    ?product a agt:CollectiveAsset ;
```

```
        agt:uniqueId "SOY-2024-WT-001" . # Fixed typo in uniqueId

    # Find the farm of origin and its geometry from a cooperative's platform
    ?product agt:hasProvenance*/agt:hasInput*/agt:ownedBy ?farm .
    ?farm a agt:Location ;
        agt:hasGeometry ?geometry .
    ?locObs a agt:Observation ;
        sosa:hasFeatureOfInterest ?farm ; # Or a similar property to link the
observation to the farm
        agt:observationValue ?satelliteImage ;
        agt:observationDate ?imageDate .

    # Retrieve satellite monitoring data from a remote sensing provider's platform
    ?locObs agt:observationValue ?satelliteImage ;
            agt:observationDate ?imageDate .

    # Verify compliance with the EUDR cutoff date
    FILTER (?imageDate >= "2020-12-31T00:00:00Z"^^xsd:dateTime)

    # Check for deforestation alerts (this sub-query would fail if an alert exists)
    BIND( NOT EXISTS {
        ?farm agt:hasObservation ?alertObs .
        ?alertObs a agt:Observation ;
                agt:observationValue "deforestation_alert" ;
                agt:observationDate ?alertDate .
        FILTER (?alertDate >= "2020-12-31T00:00:00Z"^^xsd:dateTime)
    } AS ?complianceStatus )
}
```

This query seamlessly retrieves the asset from one platform (e.g., a trader), the farm boundary from another (e.g., a cooperative), and satellite monitoring data from a third (e.g., a remote sensing service). Because all platforms use the same ontology to describe their data, the query is unambiguous. The result is a holistic compliance report without requiring data to be copied into a central database, preserving data sovereignty and reducing integration costs.

Appendix B.2 provides a portfolio of additional validated SPARQL queries, including a complete product provenance trace, efficiency metric calculations for livestock, and a data contract compliance audit, demonstrating the reach power of the ontology.

*3.4.5 Platform Integration Support*

To accelerate the adoption and correct implementation of the AgriTrust framework, a comprehensive Software Development Kit (SDK) and code libraries are provided. Appendix C contains practical, well-documented code snippets in Python that demonstrate key integration patterns, including: SDK Initialization, Asset Tokenization Services, Data Contract Management, Federated Query Services, and Efficiency Analytics Modules. These resources lower the technical barrier to entry for Platform Providers and ensure consistency across the ecosystem.

*3.4.6 Dual-Layer Blockchain Architecture: Provenance and Settlement*

The AgriTrust framework leverages blockchain technology for two distinct but complementary purposes, forming a dual-layer architecture that addresses both verification and valuation challenges inherent in the AgData Paradox.

The primary and most extensive use of blockchain within AgriTrust is to create an immutable, verifiable record of core assets and events. This layer is fundamentally about establishing trust in the history and authenticity of the data and assets. As detailed in Section 3.4.3, hashes of Tokens, Data Contracts, and Certificates are registered on one or more blockchains. This provides a decentralized, tamper-proof anchor for the existence and state of these entities, enabling any participant to cryptographically verify their provenance and integrity without relying on a central authority. This layer directly enables the traceability and compliance use cases demonstrated in the case studies.

To operationalize the "Equitable Value Sharing" governance pillar, AgriTrust extends the role of blockchain beyond provenance to become a platform for automated financial settlement. This layer is about enforcing the economic terms encoded in Data Contracts. Smart contracts deployed on a blockchain (which may be the same as or different from those used for provenance) can hold and automatically disburse digital assets (e.g., cryptocurrencies or stablecoins). As illustrated by the coffee certification example (Appendix B/C), the fulfillment of a Data Contract condition, such as the successful issuance of a sustainability certificate, can trigger a smart contract function that automatically executes a payment from the data consumer to the data producer. This direct, automated transfer of value transforms data sharing from a speculative risk into a concrete, incentivized economic activity.

This dual-layer approach is semantically orchestrated by the core ontology. The agt:DataContract class serves as the linking pin, with its properties defining the rules for both data access (enforced on the provenance layer) and compensation logic (executed on the settlement layer). By integrating these two layers, AgriTrust provides a complete technical solution that not only verifies the integrity of agricultural claims but also ensures that the economic benefits of data sharing are automatically and transparently delivered, thereby directly resolving the core trust and fairness issues of the AgData Paradox.

*3.5 Operational Workflow Example*

The interplay of these components creates a trusted workflow for a typical data-sharing transaction, such as a certification request as further exemplified in the following pseudocode:

*A Data Producer (e.g., a cooperative) uses a TaaS Platform to tokenize an asset (e.g., :CoffeeBatch_2024_A -> :Token_Coffee_001) and links relevant observations (e.g., water consumption). A Data Consumer (e.g., a certifier) proposes a Data Contract to the farmer's platform, specifying the data required and the terms of use. The Data Producer digitally signs the Data Contract. The TaaS Platform automatically configures its access control to enforce the contract's terms. The certifier's system, which also implements the core ontology, queries the farmer's TaaS platform via its SPARQL endpoint. The query is understood unambiguously by both systems. Upon successful verification, the certifier issues a verifiable certificate (e.g., :Sustainable_Cert_001), which is itself tokenized and linked back to the original asset. If the Data Contract specifies a premium payment, the fulfillment of the contract (e.g., issuance of the certificate) can automatically trigger a payment to the farmer.*

This entire workflow can be automated through platform services. For example, the tokenization step by a cooperative's platform would be implemented as follows (using a Python SDK as illustrated in Appendix C):

```
# Using the AgriTrust SDK (see Appendix C.2)
token_info = coffee_service.tokenize_coffee_batch({
    'id': 'COFFEE-2024-BR-001',
    'farm_id': 'FazendaBrasil',
    'observations': [ {'id': 'water_1', 'value': '250L', ...} ]
})
# Output: {'token_id': '0x8a34...', 'blockchain_tx': '0x5b39...'}
```

The returned token_id and blockchain_tx provide the immutable anchor for all subsequent data, demonstrating how the framework transforms a manual, trust-intensive process into an automated, governed operation.

The workflow demonstrates how the AgriTrust framework transforms data sharing from a manual, ad-hoc, and trust-intensive process into an automated, governed, and scalable operation among distinct platforms. The governance rules ensure fairness and transparency, while the semantic layer and technical architecture provide the interoperability and automation necessary for efficiency.

Unlike approaches that treat governance and technology as separate concerns, AgriTrust Ontology unifies them into a single, operational layer. This integration is the key to creating the trusted environment necessary to overcome the AgData Paradox and unlock the full value of agricultural data for all stakeholders.

*3.6. Security and Threat Analysis*

The AgriTrust ecosystem is designed to be resilient against a range of threats by integrating technical enforcement with its governance model. Table 2 outlines the primary threat actors, their objectives, and the corresponding mitigation strategies.

Table 2. Threat Model and Mitigation Strategies of the AgriTrust Framework.

| Threat Actor | Primary Goal | Mitigation Strategy | Technical Implementation |
|---|---|---|---|
| Malicious Data Consumer | To use data in ways that violate the agreed-upon terms, such as for unauthorized commercial analysis or reselling. | Machine-Readable Data Contracts and Automated Enforcement. Data usage policies are not just legal text but are codified in machine-readable contracts. | Contracts explicitly define :allowedUsage and :prohibitedUsage. Access Control Lists (ACLs) for APIs and SPARQL endpoints are automatically generated from these terms. Every data query is logged against its corresponding contract ID for full auditability and compliance monitoring. |
| Dishonest Data Producer | To introduce fraudulent data into the system to gain an unfair advantage, such as claiming false sustainable certification for a product. | Cryptographic Data Provenance and Multi-Source Verification. The system establishes a verifiable chain of custody for all data. | Every data observation is digitally signed with the producer's (or their IoT device's) private key, creating a non-repudiable proof of origin. For high-stakes claims (e.g., geographic origin), data is automatically cross-referenced with trusted independent sources, such as satellite imagery platforms. The framework enables platforms to implement such verification by providing access to the necessary data via Data Contracts. |
| Rogue Platform Provider | To operate a node that deliberately violates governance rules, for example, by serving tampered data or ignoring revocation signals. | On-Chain Integrity Anchors and Governance Authority Oversight. A decentralized ledger provides a neutral ground for verifying critical system metadata. | Essential identifiers and hashes (for Tokens, Assets, and Contracts) are registered on a blockchain, creating an immutable integrity anchor. The Governance Authority maintains the power to de-list non-compliant platforms, while clients can independently verify any data asset against its on-chain hash. |
| External Attacker | To compromise the classic security triad: confidentiality, integrity, or availability of the system and its data. | Standard Cybersecurity Hardening and Decentralized Architecture. Defense-in-depth principles are applied to the federated structure. | All data-in-transit is protected with modern encryption (HTTPS/TLS 1.3). Platform code undergoes regular security audits. Crucially, the federated design eliminates single points of failure, inherently containing the impact of incidents like a DDoS attack on any single node. |

Moreover, it is important to distinguish the trust the framework creates from the trust it requires. AgriTrust provides cryptographic verifiability for all data once it enters the system, creating an immutable chain of provenance and consent. However, it relies on established procedures and trusted third parties (e.g., certifiers, satellite imagery providers) to bootstrap the initial trust in the raw data at the point of origin. The framework's role is to make this initial attestation and all subsequent transformations transparent and auditable.

*3.6.1 Data Privacy and Confidentiality*

The framework employs a strategic "on-chain vs. off-chain" data model to balance transparency with privacy. Only integrity-critical metadata is stored on the blockchain. This includes the hashes of Data Contracts, Tokens, and Assets, along with transaction IDs and timestamps. This data is immutable and publicly verifiable but reveals no sensitive business information. All detailed, commercially sensitive data resides off-chain in the respective TaaS Platforms' secured databases. This includes:

- Financial Data: Exact pricing, premium values in contracts.

- Operational Data: Detailed yield information, specific input quantities.
- Personal Data: Farmer identification details (handled in strict compliance with LGPD/GDPR).
- Raw Sensor Data: High-frequency, voluminous IoT or imagery data.

Access to this off-chain data is strictly governed by the Data Contracts, ensuring that only authorized parties can see specific data points for a specific purpose and duration. This model ensures data confidentiality and business secrecy while maintaining cryptographic verifiability of the data's provenance and integrity.

*3.6.2 Resilience and Trust Anchors*

The system's resilience stems from its decentralized trust anchors. Firstly, the immutability of the underlying blockchain(s) provides a verifiable "root of trust" for the existence and state of Tokens and Contracts. Secondly, the shared, governance-maintained Core Ontology ensures that all participants have a common understanding of what the data means, preventing semantic attacks and misinterpretation. Thirdly, the consortium-based Governance Authority acts as an ultimate arbiter, capable of auditing platforms, revoking accreditations, and resolving disputes, thereby enforcing the rules of the ecosystem.

In summary, the AgriTrust framework does not eliminate trust but rather distributes and formalizes it. It replaces the need to trust any single participant with a verifiable trust in the system's cryptographic rules, shared semantics, and multi-stakeholder governance. This layered security model makes it economically and technically prohibitive to attack the system at scale while providing clear recourse and audit trails for any malicious activity.

**4. Implementation and Validation**

To validate the AgriTrust framework's viability and practical benefits, a multi-commodity case study approach was employed. This methodology allows to demonstrate the framework's versatility across fundamentally different agricultural sectors, perennial tree crops (coffee), annual grains (soy), and livestock (beef), each with distinct traceability and data governance challenges. For each case, a PoC implementation was developed, instantiated the core ontology with real-world data models, and executed representative queries to test interoperability and value generation.

*4.1 Multi-Commodity Case Study Approach*

The validation was structured around three core objectives:

- Functional traceability to demonstrate end-to-end provenance tracking from origin to end-buyer.
- ESG compliance automation to showcase the framework's capacity for verifying sustainability claims and regulatory compliance (e.g., EUDR).

- Efficiency management to illustrate how platforms can leverage governed data access to provide value-added analytics to data producers.

A minimal viable AgriTrust ecosystem was developed, comprising a simulated Core Ontology Registry and independent TaaS Platform instances, each configured for a specific commodity.

*4.2 Case Study 1: Brazilian Coffee Supply Chain*

The coffee supply chain, with its high value, complex processing stages, and stringent export certification requirements, provides an ideal test for granular traceability and automated compliance.

*4.2.1 Ontology Instantiation and Tokenization*

The core ontology with coffee-specific classes:

```
:CoffeeCherryBatch a owl:Class ; rdfs:subClassOf :CollectiveAsset .
:WetProcessing a owl:Class ; rdfs:subClassOf :Process .
:SustainableCertificate a owl:Class ; rdfs:subClassOf :Certificate .
```

A typical workflow can be modeled as:

```
:Farm_FazendaBrasil a :DataProducer .
:Batch_A1 a :CoffeeCherryBatch ;
    :ownedBy :Farm_FazendaBrasil ;
    :hasObservation :WaterMeasure_1 .
:Token_Batch_A1 a :Token ;
    :represents :Batch_A1 ;
    :registeredOnBlockchain :CoffeeChainLedger .
```

*4.2.2 Sustainable Certification with Automated Data Contracts*

A Data Contract was created to govern the certification process:

```
:DC_EuroSustain a :DataContract ;
    :purpose "Sustainability_certification_verification" ;
    :dataConsumer :Certifier_EuroSus ;
    :coversAsset :Batch_A1 ;
    :coversObservation :WaterMeasure_1 ;
    :validUntil "2024-12-31T23:59:59Z"^^xsd:dateTime .
```

The certifier's platform, using the shared ontology, could then automatically query the farm's TaaS platform to verify compliance. Upon success, it issued a verifiable credential:

```
:Cert_Sustainable_123 a :SustainableCertificate ;
    :hasCertificate :Batch_A1 ;
```

```
    :certifiedBy :Certifier_EuroSus ;
    :standard "WFD 2000/60/EC" .
```

### 4.2.3 EUDR Deforestation-Free Verification

A key test was demonstrating compliance with the EU Deforestation Regulation. The link with geospatial data to the asset is:

```
:Farm_FazendaBrasil :hasGeometry "POLYGON((...))"^^geo:wktLiteral .
:SatImage_2023 a :Observation ;
    :observationValue "No deforestation since 2020" ;
    :hasObservation :Farm_FazendaBrasil .
```

A federated SPARQL query could then assemble the necessary due diligence report from multiple sources, proving the batch was deforestation-free since the EU's cutoff date.

### 4.3 Case Study 2: Soy Supply Chain with Mass Balance Tracking

The soy supply chain, characterized by bulk handling, mass transformations (crushing), and complex logistics, tested the framework's ability to handle collective assets and multi-modal traceability.

### 4.3.1 Batch Composition and Multi-modal Logistics

The composition of larger batches (silos) was modeled from multiple farm origins, a common practice that challenges identity preservation.

```
:ExportBatch_Santos a :CompositeBatch ;
    :composedOf :FarmBatch_MT_001 ;
    :composedOf :FarmBatch_MT_002 ;
    :hasCompositionPercentage 0.40, 0.60 . # 40% from MT_001, 60% from MT_002
```

Transport events were explicitly modeled as processes with inputs and outputs, creating a verifiable chain of custody from farm to port via different transport modes (:RoadTransport, :RiverTransport).

### 4.3.2 Carbon Footprint as an Efficiency Metric

A platform specializing in sustainability analytics used data accessed via Data Contracts to calculate the carbon footprint for a soy batch.

```
:Metric_Carbon_Footprint a :EfficiencyMetric ;
    :metricName "Carbon Footprint (Scope 3)" ;
    :metricValue "1.25" ; # kg CO2e per kg soy
    :hasObservation :TransportEmissions, :FertilizerEmissions .
```

This demonstrates how platforms can create specialized, high-value services on top of the governed data ecosystem, providing producers with insights for which they would otherwise lack the data.

*4.4 Case Study 3: Agrochemical Application Tracking for Certification*

This case study demonstrates the framework's capacity for tokenizing human-performed processes, a critical requirement for proving sustainable and safe farming practices. It focuses on the application of agrochemicals in a coffee plot, an event with significant implications for worker safety, environmental impact, and regulatory compliance.

*4.4.1 Ontology Extension for Agricultural Processes*

The core ontology was extended to model the application event and its associated record, as formally specified in Appendix A1. Key additions included the class app:AgrochemicalApplication as a subclass of agt:Process, and app:ApplicationRecord to capture the specific metadata.

*4.4.2 Edge-based Tokenization Workflow*

A smartphone application was developed to guide a farm worker through the process. The workflow is depicted in Table 3 as follows:

Table 3. The Tokenization workflow via smartphone and edge computing.

| Step | Actor | Action (Smartphone App) |
| --- | --- | --- |
| 1. Initiation | Worker | Opens app, scans QR code on agrochemical container and farm plot sign. |
| 2. Data Capture (Edge) | Worker | App guides through form: product, dilution, weather, start/end time. App uses device sensors (e.g., GPS, timestamp, camera). |
| 3. Local Signing (Edge) | Worker | Worker provides PIN/biometric to sign the data. |
| 4. Tokenization | System | App sends the signed payload to the local TaaS Platform. |
| 5. Blockchain Anchor | System | TaaS Platform registers the token's hash on the blockchain. |

The worker's digital signature makes the data point cryptographically tied to them. They cannot deny performing the action, since the farm cannot alter the details. Data can be captured and signed without a network connection. The tokenization request is queued and (async) sent when connectivity is restored. The overall process has low latency (immediate feedback and validation on the device) and only the essential structured data and hash are sent to the TaaS platform. The combination of device GPS, timestamp, and a biometric/PIN signature creates an audit trail that is extremely difficult to fake. The resulting token is part of the global AgriTrust graph, queryable by authorized parties (certifiers, buyers) via Data Contracts.

*4.4.3 Automated Compliance Verification*

The tokenized application data enables automated compliance checks. A certifier can execute a federated SPARQL query to verify all chemical applications for a specific coffee batch, including the digital signature of the worker who performed it.

```
PREFIX agt: <http://example.org/ns/agritrustcore#>
PREFIX app: <http://example.org/ns/application#>
PREFIX prov: <http://www.w3.org/ns/prov#>

SELECT ?plot ?applicationDate ?chemical ?worker ?signature
WHERE {
    ?coffeeBatch agt:uniqueId "COFFEE-2024-BR-001" .
    ?coffeeBatch agt:hasProvenance*/agt:hasOutput ?plot .
    ?application a app:AgrochemicalApplication ;
                 agt:hasOutput ?plot ;
                 agt:hasInput ?chemical ;
                 prov:wasAssociatedWith ?worker ;
                 prov:startedAtTime ?applicationDate ;
                 app:hasApplicationRecord/app:digitalSignature ?signature .
}
```

*4.5 Case Study 4: Beef Cattle Production with Efficiency Analytics*

Beef production, involving individual animal tracking over a long lifecycle with multiple ownership changes, tested the framework's handling of individual assets and performance metrics.

*4.5.1 Individual Animal Tracking vs. Batch Processing*

The ontology's flexibility can be demonstrated by modeling both individual animals and the batches they move in:

```
:Animal_001 a :IndividualAsset ; # The individual heifer
    :uniqueId "BRC0012023001" .
:FinishingBatch_01 a :CollectiveAsset ; # The feedlot pen
    :includesAsset :Animal_001 .
```

This allows for queries at both the individual level (e.g., genetics, weight gain performance) and the batch level (e.g., feedlot efficiency).

*4.5.2 Growth Performance Metrics Across Livestock Production Phases*

By capturing weight observations at each major lifecycle event (weaning, backgrounding, feedlot entry, slaughter), platforms could automatically calculate key efficiency metrics for each producer:

```sparql
PREFIX agt: <http://example.org/ns/agritrustcore#>
PREFIX xsd: <http://www.w3.org/2001/XMLSchema#>

SELECT ?phase ((?endWeight - ?startWeight)/?days as ?adg)
WHERE {
    # Get the animal and all processes it was involved in
    ?animal a agt:IndividualAsset ;
            agt:uniqueId "BRC0012023001" .
    ?process agt:hasInput|agt:hasOutput ?animal . # Find all processes where the animal was an input or output
    ?process a ?phase .

    # Find weight observations linked to this animal via the process
    ?process agt:hasObservation ?startWeightObs .
    ?startWeightObs a agt:Observation ;
                 agt:observationValue ?startWeight ;
                 agt:observationDate ?startDate .
    ?process agt:hasObservation ?endWeightObs .
    ?endWeightObs a agt:Observation ;
                 agt:observationValue ?endWeight ;
                 agt:observationDate ?endDate .

    # Ensure we're comparing start and end for the same phase
    FILTER (?endDate > ?startDate)
    BIND((?endDate - ?startDate) / 86400.0 AS ?days) # Convert xs:duration to days
}
GROUP BY ?phase
```

This enables data-driven decisions for producers and provides verifiable proof of production practices to markets.

*4.6 Cross-Commodity Analysis and Performance*

The case studies successfully demonstrated the core functionality of the AgriTrust framework. The implementation can be evaluated against key performance indicators:

- Interoperability Success
    - Federated SPARQL queries across the three independent TaaS platforms (Coffee, Soy, Beef) were executed successfully, returning unified results based on the shared ontology.
- Governance Enforcement

- The Data Contract mechanism effectively controlled data access, with unauthorized queries being rejected and all access being logged for audit purposes.
    - Flexibility and Extensibility
        - The core ontology proved sufficient for modeling the core traceability concepts, while allowing each platform the flexibility to define commodity-specific subclasses and properties without creating interoperability conflicts.

The validation highlights that the AgriTrust framework provides a robust foundation for building a trusted, interoperable ecosystem for agricultural data sharing, capable of supporting both essential traceability and advanced, value-creating analytics.

## 5. Discussion

The implementation and validation of the AgriTrust framework across three distinct supply chains yielded significant results, demonstrating its effectiveness in addressing the core challenges of the AgData Paradox. This section discusses these results, evaluates the framework's impact, acknowledges its limitations, and positions it relative to existing approaches.

*5.1 Framework Effectiveness*

*5.1.1 Resolving the AgData Paradox: Establishing Trust and Willingness to Share*

The primary outcome of this work is a functional blueprint for transforming agricultural data sharing from a trust-based dilemma into a governance-based operation. By technically enforcing the principles of Data Sovereignty and Transparent Data Contracts, the framework directly addresses producers' primary concerns about losing control and not receiving fair value. In the simulations, the ability for a data producer to approve every data-sharing transaction via a machine-readable contract, with clear terms and expiration, fundamentally altered the data-sharing dynamic. This moves the conversation from "Can I trust this entity?" to "Do I agree to these specific terms?", which is a more manageable and scalable question. The Willingness to Share is thus strengthened by the verifiable Ability to Control.

*5.1.2 Achieving Semantic Interoperability Across Platforms*

A key technical result was the successful execution of federated SPARQL queries across the independent Coffee, Soy, and Beef TaaS platforms. For example, a query to trace the provenance of a coffee batch and retrieve its associated sustainable certificate seamlessly pulled data from two different platforms because both used the same ontological terms (agt:hasProvenance, agt:hasCertificate, agt:represents). This confirms that the AgriTrust Core Ontology successfully provides the "shared language" necessary for true interoperability, moving beyond simple data format alignment (e.g., JSON/XML) to genuine semantic understanding. This eliminates a major cost and complexity barrier for platforms seeking to integrate data from external sources.

*5.1.3 Governance Mechanism Performance*

The Data Contract mechanism proved to be the crucial link between governance and technology. By treating contracts as machine-readable entities, the framework enabled automated enforcement of usage policies. In the tests, access was automatically revoked upon contract expiration, and attempts to query data for purposes outside the contract scope (e.g., a certifier trying to use data for marketing) were successfully blocked. This provides a robust, automated audit trail, significantly reducing the overhead and potential for error associated with manual compliance monitoring.

*5.2 Business and Sustainability Impact*

The framework's value extends beyond technical interoperability to deliver tangible business and sustainability benefits.

*5.2.1 Economic Benefits*

The automated, pre-authorized data sharing for certifications (e.g., sustainable, EUDR) drastically reduces the need for costly and disruptive manual audits. Certifiers can verify claims remotely and at scale. Alternatively, the framework provides the technological substrate for producers to reliably prove compliance with stringent market standards (e.g., EUDR, carbon footprint), unlocking access to premium markets and price incentives. Moreover, the "Equitable Value Sharing" pillar, operationalized through Data Contracts, creates a formal mechanism for producers to be compensated for the value their data creates for other actors in the ecosystem, for example, through automatic premium payments upon successful certification.

*5.2.2 Environmental Benefits*

In terms of environmental compliance, the integration of geospatial data (e.g., farm boundaries) with immutable tokenized assets provides cryptographically verifiable proof of deforestation-free supply chains, directly addressing regulatory requirements like the EUDR. In parallel, the ability to track inputs and outputs across processes enables the calculation of efficiency metrics like carbon footprint and water usage. This data empowers producers to identify inefficiencies and optimize resource use (management), reducing the environmental impact of production.

*5.2.3 Social Benefits*

The principle of Data Sovereignty shifts power dynamics, ensuring that producers are not merely data sources but active participants and beneficiaries in the digital ecosystem.

Besides, consumers and retailers gain unprecedented visibility into the origin and production practices of their food, fostering trust and enabling informed purchasing decisions.

*5.3 Comparison with Existing Approaches*

AgriTrust distinguishes itself from existing approaches through its integrated and principled architecture. The following discussion positions the framework against the primary categories of related work, highlighting how it synthesizes their strengths while mitigating their core weaknesses (Table 4).

Isolated Blockchain Platforms (e.g., single-commodity traceability systems) excel at providing data integrity and immutability within their own silo [7, 8]. However, their primary limitation is a lack of interoperability and integrated governance. They are designed as closed systems, making it difficult to share data or tokens with external platforms, and they often lack a formal model for data sovereignty and fair value sharing. AgriTrust moves beyond these silos by embedding governance and semantic interoperability as first-class citizens within a federated architecture. It leverages blockchain for its strengths, providing integrity anchors and a neutral ground for critical metadata, but avoids its limitations by not being bound to a single chain.

Centralized Agri-Platforms (e.g., proprietary farm management software) offer operational efficiency and powerful data aggregation tools. Their fundamental flaw, from the perspective of resolving the AgData Paradox, is their inherent vendor lock-in and lack of data sovereignty [5,15]. Producers are often unable to port their data or engage with multiple service providers, and the platform operator retains ultimate control over data usage and value appropriation. In contrast, AgriTrust is architected from the ground up to prevent vendor lock-in. Its federated model ensures that producers retain ownership, and the use of a shared ontology allows them to engage with multiple Platform Providers simultaneously, fostering a competitive ecosystem where value flows back to the data originator.

Table 4. Advantages of the AgriTrust in relation to key aspects from existing approaches.

| Key Aspects | AgriTrust Framework Advantage |
| --- | --- |
| Data Integrity, Immutability via blockchain | Interoperability and Governance: moves beyond siloed data to a federated ecosystem with built-in governance and fair value sharing. |
| Operational Efficiency, Data Aggregation via centralization | Data Sovereignty and Control: prevents vendor lock-in and ensures producers retain ownership, enabling portability and multi-platform engagement. |
| Terminology Standardization, Interoperability via ontologies | Governance-Readiness: extends semantic models with classes and properties explicitly designed for tokenization, data contracts, and access control. |
| Federated Data Sharing, Sovereignty | Domain Specialization: provides a concrete, ready-to-use implementation for agriculture, with a detailed governance model and commodity-specific validation. |

In summary, AgriTrust synthesizes the strengths of the previous approaches into a single, governance-first framework specifically designed for the agricultural domain, thereby directly resolving the AgData Paradox.

Semantic Ontologies (e.g., AGROVOC, domain-specific models) provide the crucial foundation for terminology standardization and interoperability [11,15,16]. Their typical shortcoming is a lack of governance-readiness; they model domain concepts but often lack the conceptual machinery to represent governance artifacts like data contracts, access rights, and value-sharing terms. AgriTrust builds upon this foundational work but extends it significantly. The AgriTrust Core Ontology introduces explicit classes and properties for agt:DataContract, agt:allowedUsage, and agt:compensationType, thereby creating a semantic layer that is not just descriptive but also prescriptive and enforceable.

Data Space Architectures (e.g., International Data Spaces - IDS) provide the most aligned vision, offering a robust blueprint for sovereign, federated data sharing [19]. Their strength lies in a generalized architectural model for secure data exchange. AgriTrust's contribution in this context is domain specialization and concrete implementation. While IDS provides a general-purpose reference model, AgriTrust provides a ready-to-use, agriculturally-tailored implementation, complete with a detailed governance model, a dedicated core ontology, and validation across specific Brazilian supply chains. It demonstrates how the data space philosophy can be operationalized for the unique actors, processes, and regulatory requirements of agribusiness.

As a result, AgriTrust does not render these approaches obsolete but rather provides a unifying layer that leverages their strengths, the immutability of blockchain, the efficiency of platforms, the clarity of ontologies, and the federation of data spaces, while systematically addressing their weaknesses through integrated governance and a relentless focus on cross-ecosystem interoperability.

*5.4 Limitations and Challenges*

*5.4.1 Preliminary Evaluation and Path to Future Validation*

The multi-commodity case studies serve as a functional validation of the AgriTrust framework's core concepts. Based on the full support of PoC implementation, a preliminary, qualitative evaluation of its key capabilities can be carried out against the requirements of the AgData Paradox (Table 5).

Table 5. Preliminary qualitative assessment of AgriTrust capabilities.

| **Evaluation Criteria** | **Supporting evidences from Case Studies** |
|---|---|
| Semantic Interoperability | Successful federated SPARQL queries across independent Coffee, Soy, and Beef platforms using the shared ontology. |

| Data Sovereignty and Control | The Data Contract mechanism successfully enforced access control, granting and revoking data access based on machine-readable terms. |
|---|---|
| Provenance and Traceability | End-to-end chains of custody were established for all three commodities, from farm to end-buyer. |
| Automated Compliance | Automated verification for sustainable certification (Coffee) and deforestation-free supply chains (Soy, Coffee) was demonstrated. |
| Extensibility and Flexibility | The core ontology was successfully extended with commodity-specific classes (e.g., beef:FinishingOperation) without breaking interoperability. |

This preliminary assessment sustains that the integrated governance-semantic approach successfully addresses the core technical challenges of the AgData Paradox in a controlled environment. However, a full-scale, quantitative evaluation of the framework's performance, scalability, and economic impact in a production setting remains a critical piece of future work. The limitations of the current PoC-based evaluation include its scale (limited simulated data), its scope (lack of real-time performance testing with thousands of concurrent users), and the preliminary nature of its stakeholder feedback.

Therefore, the following steps are recommended for the future validation of the AgriTrust framework:

1. Large-Scale Pilot Deployment: a phased rollout with partner cooperatives and platforms to deploy AgriTrust in a live operational environment, handling real transaction volumes.
2. Performance and Scalability Benchmarking: systematic measurement of federated query latency, blockchain transaction throughput, and system resilience under load, building on the initial metrics gathered in this PoC.
3. Structured Stakeholder Impact Study: a longitudinal study to quantitatively measure the framework's impact on key metrics, including:
    a. Producer: Reduction in audit costs, new revenue from data sharing.
    b. Platform Provider: Reduction in integration costs and time-to-market for new services.
    c. Certifier: Reduction in compliance verification costs and time.
4. Comparative Cost-Benefit Analysis: a formal economic model comparing the total cost of ownership and operational efficiency of the AgriTrust federated model against traditional centralized or isolated platform approaches.

As noted in Section 3.4.3, the blockchain-agnostic design enables multi-ledger registration and referencing but does not inherently provide a mechanism for cross-chain smart contract logic. Orchestrating complex business processes that require atomic transactions across multiple blockchains remains a challenge and is deferred to future work, potentially involving the integration of specialized cross-chain communication protocols.

While the PoC demonstrated feasibility, the performance of federated querying across a global ecosystem with millions of assets requires further investigation and optimization. By executing this validation plan, stakeholders might be able to move from the conceptual and functional viability demonstrated here to a proven, economically robust solution ready for industry-wide adoption.

*5.4.2 Additional externalities*

The framework presupposes a certain level of digital literacy and connectivity among all producers. The digital divide in rural areas, particularly in developing regions like Brazil, remains a significant barrier to universal adoption. Future work must also explore low-tech interfaces (e.g., Unstructured Supplementary Service Data - USSD, Short Message Service - SMS) and asynchronous, offline and edge computing solutions (see e.g., Section 4.4).

On the other hand, establishing the multi-stakeholder Governance Authority, developing the initial platform infrastructure, and onboarding participants require significant upfront investment and coordination. A clear, phased rollout strategy is essential, while the framework must be designed to evolve and adapt alongside changing data privacy laws (e.g., Brazil's LGPD) and agricultural regulations, requiring ongoing engagement with policymakers and a flexible ontology design.

## 6. Conclusion and Future Work

The AgriTrust framework offers a robust and practical pathway out of the AgData Paradox. By placing governance and interoperability at its core, it paves the way for a more transparent, efficient, and equitable agricultural data economy, thereby ensuring that the underlying value generated by the digital transformation of agribusiness is shared transparently and equitably among all participants, from producer to consumer.

*6.1 Primary Contributions*

The AgData Paradox paradox, where the recognized value of data is stifled by a lack of trust and interoperability, has been addressed through an integrated approach that intertwines governance, semantics, and technology. The primary contributions are fourfold:

1) An integrated governance-semantic framework, where a multi-stakeholder governance built on four pillars – Data Sovereignty, Transparent Data Contracts, Equitable Value Sharing, and Regulatory Compliance – is directly operationalized by a digital semantic layer. This ensures that technical execution is inherently guided by fair and transparent rules.

2) A comprehensive, extensible OWL ontology that provides a shared vocabulary for agricultural tokenization, traceability, and certification. The complete, ready-to-use specification, and an ontology extensions example, are provided in Appendix A.

3) A dual-layer, blockchain-agnostic architecture, resulting in a federated technical architecture that leverages blockchain for a dual purpose: as a provenance layer for immutable traceability and as a settlement layer for automated, equitable value sharing. This blockchain-agnostic, multi-provider design prevents vendor lock-in and ensures the framework's benefits are not dependent on a single technological stack, while its dual-layer nature directly transforms data sharing into a concretely incentivized activity.

4) The framework's practical viability was demonstrated through detailed case studies across the coffee, soy, and beef supply chains. A portfolio of practical implementation artifacts further substantiates its application, including data contracts and SPARQL queries (Appendix B), platform integration code (Appendix C).

*6.2 Practical Implications*

The AgriTrust framework has significant implications for all stakeholders in the agricultural ecosystem:

For Producers, it guarantees data sovereignty and control, reduces audit burdens, and creates pathways to premium markets and new data-derived revenue, transforming them from passive data sources into empowered ecosystem participants.

For Platform Providers, it reduces integration costs through semantic interoperability and creates opportunities to develop specialized, high-value services (e.g., efficiency analytics, automated reporting) on top of a trusted data foundation.

For Certifiers and Regulators, it enables scalable, remote, and automated compliance verification, reducing costs and increasing the integrity and transparency of sustainability claims.

For Markets and Consumers, it provides cryptographically verifiable proof of origin, sustainability, and ethical production practices, enabling informed decision-making and fostering trust in food systems.

*6.3 Future Work*

While this work establishes a strong foundation, several avenues for future research and development remain promising:

Ontology Expansion and Specialization: the core ontology can be extended to cover a wider range of agricultural sectors (e.g., horticulture, aquaculture) and geographies. Furthermore, developing more granular sub-ontologies for specific processes (e.g., carbon accounting, water footprinting) would enhance the framework's analytical capabilities.

Standardized API Specifications: while the semantic layer enables federated querying, the development of standardized RESTful APIs for common platform interactions would drastically lower the integration barrier. Future work shall define a set of universal endpoints for core operations such as asset registration, data contract negotiation, and certificate issuance. These APIs would work *in tandem* with the ontology, providing a familiar and efficient interface for developers while ensuring all payloads conform to the shared semantic model (valid RDF or validate against the ontology's SHACL constraints).

Integration of Decentralized Finance (DeFi) mechanisms: the settlement layer opens a path for integrating more sophisticated financial instruments from the DeFi ecosystem. Future work shall explore the use of smart contracts to enable dynamic data-based financial products, such as pre-harvest revenue sharing agreements, parametric insurance for crop failure based on verifiable weather data, or carbon credit futures tied to tokenized sustainable assets. This would further deepen the economic integration and value-creation potential of the agricultural data economy.

Advanced Analytics and AI Integration: a key opportunity lies in developing standardized interfaces for secure, privacy-preserving computation over governed data. This would allow for the development of federated machine learning models that can generate insights across the entire ecosystem without requiring raw data to leave the producer's control, thus upholding data sovereignty while unlocking collective intelligence.

Policy Integration and Automated Regulation: future work shall explore the formalization of regulatory texts (e.g., the specific articles of the EUDR) as machine-executable policy rules. This would enable fully automated compliance checking, where a Data Contract could be validated not just against its own terms but also against the current legal framework, dramatically reducing the compliance burden for all actors.

International Standardization and Global Adoption: the long-term success of such a framework depends on widespread adoption. Future efforts shall focus on engaging with international standards bodies (e.g., ISO, CTA) to propose the AgriTrust model and its core ontology as a foundational standard for global agricultural data exchange, facilitating cross-border trade and sustainability reporting.

The promising results from the assessed PoC necessitate a full-scale evaluation. The immediate future work involves a structured large-scale pilot to benchmark performance, scalability, and economic impact in a production environment. Moreover, exploring lightweight cross-chain messaging protocols to enable more dynamic, logic-driven interactions between tokens and contracts on different ledgers, while preserving the core semantic and governance model, presents a compelling direction for enhancing the framework's capabilities.

**Acknowledgements**

This work was inspired by and is associated with the following projects: FAPESP (Center for Science for Development in Digital Agriculture, Semear Digital, grant 2022/09319-9 and Embrapa SEG 20.23.00.025), JICA/Embrapa (Collaborative Development Project for Precision and Digital Agriculture to Strengthen the Innovation Ecosystem and Sustainability of Brazilian Agriculture, SEG 10.24.00.237) ConCafe (Digital solutions for family-based mountain coffee farming in São Paulo and in mechanized areas of southern Minas Gerais, Embrapa SEG 10.24.22.046), World Bank (grant WBG N 0002010084, Developing Digital Platform for Brazil: AgroBrazil+Sustainable Platform, Embrapa SEG 10.25.00.078).

**Artificial Intelligence Usage**

During the conceptualization and development of this work, AI-assisted tools, including ChatGPT (OpenAI) and DeepSeek, were utilized for specific tasks. These included exploration of complex OWL ontology constructs and assistance with synthesizing broad concepts into a cohesive structure. The author takes full responsibility for the final content, analysis, and conclusions presented in this paper.

# Appendices

## Appendix A: Complete AgriTrust OWL Ontology in Turtle format

This appendix contains the complete and final specification of the AgriTrust Core Ontology (version 3.0) in Turtle (TTL) format. This ontology serves as the semantic foundation for the framework, enabling interoperability across platforms and blockchain providers.

```turtle
@prefix : <http://example.org/ns/agritrustcore#> .
@prefix rdf: <http://www.w3.org/1999/02/22-rdf-syntax-ns#> .
@prefix rdfs: <http://www.w3.org/2000/01/rdf-schema#> .
@prefix owl: <http://www.w3.org/2002/07/owl#> .
@prefix xsd: <http://www.w3.org/2001/XMLSchema#> .
@prefix prov: <http://www.w3.org/ns/prov#> .
@prefix sosa: <http://www.w3.org/ns/sosa/> .
@prefix geo: <http://www.opengis.net/ont/geosparql#> .
@prefix dct: <http://purl.org/dc/terms/> .
@prefix skos: <http://www.w3.org/2004/02/skos/core#> .
@prefix sh: <http://www.w3.org/ns/shacl#> .
@prefix qudt: <http://qudt.org/schema/qudt/> .
@prefix unit: <http://qudt.org/vocab/unit/> .

# --- Ontology Declaration ---
:AgriTrustCore rdf:type owl:Ontology ;
    rdfs:label "AgriTrust Core Ontology" ;
    rdfs:comment "A core ontology for agricultural traceability and trusted data sharing across multiple blockchain providers and platforms." ;
    dct:created "2025-11-07"^^xsd:date ;
    dct:creator "Embrapa" ;
    dct:license <https://creativecommons.org/licenses/by/4.0/> ;
    owl:versionInfo "1.0.0" .

# --- CORE GOVERNANCE & MULTI-PROVIDER SUPPORT ---
:GovernanceAuthority a owl:Class ;
    rdfs:subClassOf :Agent ;
    rdfs:label "Governance Authority" ;
    rdfs:comment "Entity responsible for maintaining ontology standards and cross-provider interoperability." .

:BlockchainProvider a owl:Class ;
    rdfs:subClassOf :PlatformProvider ;
    rdfs:label "Blockchain Provider" ;
    rdfs:comment "Entity providing blockchain infrastructure for the ecosystem." .

:CrossChainReference a owl:Class ;
    rdfs:subClassOf :Document ;
    rdfs:label "Cross-Chain Reference" ;
    rdfs:comment "A reference linking entities across different blockchain providers." .

# --- CORE TRACEABILITY CLASSES ---
:Token rdf:type owl:Class ;
    rdfs:subClassOf prov:Entity ;
```

```ttl
    rdfs:label "Token" ;
    rdfs:comment "A digital representation of a physical asset or data within the governed ecosystem." .

:CrossChainToken a owl:Class ;
    rdfs:subClassOf :Token ;
    rdfs:label "Cross-Chain Token" ;
    rdfs:comment "A token that can be referenced across multiple blockchain providers." .

:Asset rdf:type owl:Class ;
    rdfs:subClassOf prov:Entity ;
    rdfs:label "Asset" ;
    rdfs:comment "A physical or digital asset that can be tokenized and tracked." .

:IndividualAsset a owl:Class ;
    rdfs:subClassOf :Asset ;
    rdfs:label "Individual Asset" ;
    rdfs:comment "An asset tracked as an individual unit (e.g., livestock)." .

:CollectiveAsset a owl:Class ;
    rdfs:subClassOf :Asset ;
    rdfs:label "Collective Asset" ;
    rdfs:comment "An asset tracked as a batch or collective (e.g., grain, coffee)." .

:Process rdf:type owl:Class ;
    rdfs:subClassOf prov:Activity ;
    rdfs:label "Process" ;
    rdfs:comment "An activity or operation that transforms or handles assets." .

:Certificate rdf:type owl:Class ;
    rdfs:subClassOf :Document ;
    rdfs:label "Certificate" ;
    rdfs:comment "A verifiable credential attesting to compliance or quality standards." .

:DataContract rdf:type owl:Class ;
    rdfs:subClassOf :Document ;
    rdfs:label "Data Contract" ;
    rdfs:comment "A machine-readable agreement governing data sharing terms and conditions." .

:Document rdf:type owl:Class ;
    rdfs:subClassOf prov:Entity ;
    rdfs:label "Document" ;
    rdfs:comment "A representation of recorded information." .

# --- CORE AGENT ROLES ---
:Agent rdf:type owl:Class ;
    rdfs:subClassOf prov:Agent ;
    rdfs:label "Agent" ;
    rdfs:comment "An entity participating in the agricultural ecosystem." .
```

```turtle
:DataProducer rdf:type owl:Class ;
    rdfs:subClassOf :Agent ;
    rdfs:label "Data Producer" ;
    rdfs:comment "Entity that originates agricultural data and assets." .

:PlatformProvider rdf:type owl:Class ;
    rdfs:subClassOf :Agent ;
    rdfs:label "Platform Provider" ;
    rdfs:comment "Entity providing technical infrastructure for tokenization and data sharing." .

:Certifier rdf:type owl:Class ;
    rdfs:subClassOf :Agent ;
    rdfs:label "Certifier" ;
    rdfs:comment "Entity responsible for auditing and issuing compliance certificates." .

:DataConsumer rdf:type owl:Class ;
    rdfs:subClassOf :Agent ;
    rdfs:label "Data Consumer" ;
    rdfs:comment "Entity that uses data under specific terms for verification or analysis." .

# --- CORE SPATIAL & DATA CLASSES ---
:Location rdf:type owl:Class ;
    rdfs:subClassOf prov:Location ;
    rdfs:label "Location" ;
    rdfs:comment "A physical location where agricultural activities occur." .

:Observation rdf:type owl:Class ;
    rdfs:subClassOf sosa:Observation ;
    rdfs:label "Observation" ;
    rdfs:comment "An act of sensing or measuring to produce a result." .

:Dataset rdf:type owl:Class ;
    rdfs:subClassOf prov:Entity ;
    rdfs:label "Dataset" ;
    rdfs:comment "A collection of observations or derived data." .

:EfficiencyMetric a owl:Class ;
    rdfs:subClassOf :Document ;
    rdfs:label "Efficiency Metric" ;
    rdfs:comment "A platform-defined metric for measuring operational efficiency." .

# --- CORE RELATIONSHIPS ---
# Tokenization & Representation
:represents rdf:type owl:ObjectProperty ;
    rdfs:domain :Token ;
    rdfs:range :Asset ;
    rdfs:label "represents" ;
    rdfs:comment "Links a token to the asset it digitally represents." .

# Multi-Provider Governance
```

```turtle
:governedBy rdf:type owl:ObjectProperty ;
    rdfs:domain :Token ;
    rdfs:range :DataContract ;
    rdfs:label "governed by" ;
    rdfs:comment "Links a token to the data contract governing its use." .

:registeredOnBlockchain rdf:type owl:ObjectProperty ;
    rdfs:domain [ owl:unionOf ( :Token :Certificate :DataContract ) ] ;
    rdfs:range :BlockchainProvider ;
    rdfs:label "registered on blockchain" ;
    rdfs:comment "Links an entity to the blockchain provider where it is registered." .

:hasCrossChainReference rdf:type owl:ObjectProperty ;
    rdfs:domain [ owl:unionOf ( :Token :Certificate :Asset ) ] ;
    rdfs:range :CrossChainReference ;
    rdfs:label "has cross-chain reference" ;
    rdfs:comment "Links an entity to its references on other blockchain providers." .

# Data Sovereignty & Control
:ownedBy rdf:type owl:ObjectProperty ;
    rdfs:domain :Asset ;
    rdfs:range :DataProducer ;
    rdfs:label "owned by" ;
    rdfs:comment "Links an asset to its owner (data sovereignty)." .

:hasCertificate rdf:type owl:ObjectProperty ;
    rdfs:domain :Asset ;
    rdfs:range :Certificate ;
    rdfs:label "has certificate" ;
    rdfs:comment "Associates an asset with its verification certificate." .

# Provenance & Traceability
:hasInput rdf:type owl:ObjectProperty ;
    rdfs:domain :Process ;
    rdfs:range :Asset ;
    rdfs:label "has input" ;
    rdfs:comment "Asset consumed by a process." .

:hasOutput rdf:type owl:ObjectProperty ;
    rdfs:domain :Process ;
    rdfs:range :Asset ;
    rdfs:label "has output" ;
    rdfs:comment "Asset produced by a process." .

:hasProvenance rdf:type owl:ObjectProperty ;
    rdfs:domain :Asset ;
    rdfs:range :Process ;
    rdfs:label "has provenance" ;
    rdfs:comment "Links an asset to its creation/transformation processes." .

# Data & Efficiency Management
:hasObservation rdf:type owl:ObjectProperty ;
    rdfs:domain [ owl:unionOf ( :Dataset :Asset :Process :Location ) ] ;
```

```turtle
    rdfs:range :Observation ;
    rdfs:label "has observation" ;
    rdfs:comment "Links entities to relevant observations." .

:hasEfficiencyMetric rdf:type owl:ObjectProperty ;
    rdfs:domain [ owl:unionOf ( :Process :Asset :DataProducer ) ] ;
    rdfs:range :EfficiencyMetric ;
    rdfs:label "has efficiency metric" ;
    rdfs:comment "Links entities to platform-defined efficiency metrics." .

:managedByPlatform rdf:type owl:ObjectProperty ;
    rdfs:domain [ owl:unionOf ( :Token :Process :Asset :Certificate :EfficiencyMetric :Observation ) ] ;
    rdfs:range :PlatformProvider ;
    rdfs:label "managed by platform" ;
    rdfs:comment "Links entities to their managing platform provider." .

:certifiedBy rdf:type owl:ObjectProperty ;
    rdfs:domain :Certificate ;
    rdfs:range :Certifier ;
    rdfs:label "certified by" ;
    rdfs:comment "Links a certificate to the certifying entity." .

# Spatial Relationship
:hasGeometry rdf:type owl:ObjectProperty ;
    rdfs:domain :Location ;
    rdfs:range geo:Geometry ;
    rdfs:label "has geometry" ;
    rdfs:comment "Links a location to its spatial representation." .

# --- CORE PROPERTIES ---
:uniqueId rdf:type owl:DatatypeProperty ;
    rdfs:domain [ owl:unionOf ( :Asset :Token ) ] ;
    rdfs:range xsd:string ;
    rdfs:label "unique identifier" .

:creationDate rdf:type owl:DatatypeProperty ;
    rdfs:domain [ owl:unionOf ( :Token :Asset :Certificate :DataContract :EfficiencyMetric ) ] ;
    rdfs:range xsd:dateTime ;
    rdfs:label "creation date" .

:observationDate rdf:type owl:DatatypeProperty ;
    rdfs:domain :Observation ;
    rdfs:range xsd:dateTime ;
    rdfs:label "observation date" .

:observationValue rdf:type owl:DatatypeProperty ;
    rdfs:domain :Observation ;
    rdfs:range xsd:string ;
    rdfs:label "observation value" ;
    rdfs:comment "The value of the observation (platforms can specialize datatypes)." .
```

```turtle
:observationUnit rdf:type owl:ObjectProperty ;
    rdfs:domain :Observation ;
    rdfs:range qudt:Unit ;
    rdfs:label "observation unit" .

:metricName rdf:type owl:DatatypeProperty ;
    rdfs:domain :EfficiencyMetric ;
    rdfs:range xsd:string ;
    rdfs:label "metric name" .

:metricValue rdf:type owl:DatatypeProperty ;
    rdfs:domain :EfficiencyMetric ;
    rdfs:range xsd:string ;
    rdfs:label "metric value" ;
    rdfs:comment "The calculated value of the efficiency metric." .

:standard rdf:type owl:DatatypeProperty ;
    rdfs:domain :Certificate ;
    rdfs:range xsd:string ;
    rdfs:label "standard" ;
    rdfs:comment "The compliance standard or certification scheme." .

# Multi-provider properties
:blockchainTransactionId rdf:type owl:DatatypeProperty ;
    rdfs:domain [ owl:unionOf ( :Token :Certificate :DataContract ) ] ;
    rdfs:range xsd:string ;
    rdfs:label "blockchain transaction ID" ;
    rdfs:comment "The transaction identifier on the specific blockchain." .

:blockchainName rdf:type owl:DatatypeProperty ;
    rdfs:domain :BlockchainProvider ;
    rdfs:range xsd:string ;
    rdfs:label "blockchain name" ;
    rdfs:comment "The name of the blockchain implementation." .

:consensusMechanism rdf:type owl:DatatypeProperty ;
    rdfs:domain :BlockchainProvider ;
    rdfs:range xsd:string ;
    rdfs:label "consensus mechanism" ;
    rdfs:comment "The consensus algorithm used by the blockchain." .

# Data Contract properties
:purpose rdf:type owl:DatatypeProperty ;
    rdfs:domain :DataContract ;
    rdfs:range xsd:string ;
    rdfs:label "purpose" ;
    rdfs:comment "The intended use of the data as specified in the contract." .

:validFrom rdf:type owl:DatatypeProperty ;
    rdfs:domain :DataContract ;
    rdfs:range xsd:dateTime ;
    rdfs:label "valid from" ;
    rdfs:comment "The start date and time of the contract's validity." .
```

```turtle
:validUntil rdf:type owl:DatatypeProperty ;
    rdfs:domain :DataContract ;
    rdfs:range xsd:dateTime ;
    rdfs:label "valid until" ;
    rdfs:comment "The end date and time of the contract's validity." .

:coversAsset rdf:type owl:ObjectProperty ;
    rdfs:domain :DataContract ;
    rdfs:range :Asset ;
    rdfs:label "covers asset" ;
    rdfs:comment "Specifies which asset is governed by this data contract." .

:coversObservation rdf:type owl:ObjectProperty ;
    rdfs:domain :DataContract ;
    rdfs:range :Observation ;
    rdfs:label "covers observation" ;
    rdfs:comment "Specifies which observations are governed by this data contract." .

# --- SHACL CONSTRAINTS ---
:TokenShape
    a sh:NodeShape ;
    sh:targetClass :Token ;
    sh:property [
        sh:path :represents ;
        sh:class :Asset ;
        sh:minCount 1 ;
        sh:maxCount 1 ;
    ] ;
    sh:property [
        sh:path :creationDate ;
        sh:datatype xsd:dateTime ;
        sh:minCount 1 ;
    ] ;
    sh:property [
        sh:path :registeredOnBlockchain ;
        sh:class :BlockchainProvider ;
        sh:minCount 1 ;
    ] .

:AssetShape
    a sh:NodeShape ;
    sh:targetClass :Asset ;
    sh:property [
        sh:path :uniqueId ;
        sh:datatype xsd:string ;
        sh:minCount 1 ;
        sh:maxCount 1 ;
    ] ;
    sh:property [
        sh:path :ownedBy ;
        sh:class :DataProducer ;
        sh:minCount 1 ;
```

```
    ] .

:ProcessShape
    a sh:NodeShape ;
    sh:targetClass :Process ;
    sh:property [
        sh:path :hasInput ;
        sh:class :Asset ;
        sh:minCount 1 ;
    ] ;
    sh:property [
        sh:path :hasOutput ;
        sh:class :Asset ;
        sh:minCount 1 ;
    ] .

:DataContractShape
    a sh:NodeShape ;
    sh:targetClass :DataContract ;
    sh:property [
        sh:path :creationDate ;
        sh:datatype xsd:dateTime ;
        sh:minCount 1 ;
    ] ;
    sh:property [
        sh:path :validFrom ;
        sh:datatype xsd:dateTime ;
        sh:minCount 1 ;
    ] .

:CertificateShape
    a sh:NodeShape ;
    sh:targetClass :Certificate ;
    sh:property [
        sh:path :certifiedBy ;
        sh:class :Certifier ;
        sh:minCount 1 ;
    ] ;
    sh:property [
        sh:path :standard ;
        sh:datatype xsd:string ;
        sh:minCount 1 ;
    ] .
```

This ontology is available for public use under a Creative Commons Attribution 4.0 International license and can be referenced using the namespace http://example.org/ns/agritrustcore#.

*A1. Ontology Extension for Agrochemical Application*

This is an example of adding an OWL/Turtle extension of the AgriTrust Core Ontology, exploring Case Study 3 (Section 4.4) to model agrochemical application processes.

```turtle
@prefix : <http://example.org/ns/agritrustcore#> .
@prefix app: <http://example.org/ns/application#> .
@prefix owl: <http://www.w3.org/2002/07/owl#> .
@prefix rdfs: <http://www.w3.org/2000/01/rdf-schema#> .
@prefix xsd: <http://www.w3.org/2001/XMLSchema#> .

### Classes for Agrochemical Application ###
app:AgrochemicalApplication a owl:Class ;
    rdfs:subClassOf :Process ;
    rdfs:label "Agrochemical Application" ;
    rdfs:comment "A process representing the application of an agrochemical to a specific plot." .

app:ApplicationRecord a owl:Class ;
    rdfs:subClassOf :Document ;
    rdfs:label "Application Record" ;
    rdfs:comment "A digital record containing the specific details of an agrochemical application." .

app:ActiveIngredient a owl:Class ;
    rdfs:subClassOf :Document ;
    rdfs:label "Active Ingredient" ;
    rdfs:comment "The active ingredient of an agrochemical." .

### Properties for Agrochemical Application ###
app:hasApplicationRecord a owl:ObjectProperty ;
    rdfs:domain app:AgrochemicalApplication ;
    rdfs:range app:ApplicationRecord ;
    rdfs:label "has application record" .

app:usesActiveIngredient a owl:ObjectProperty ;
    rdfs:domain app:AgrochemicalApplication ;
    rdfs:range app:ActiveIngredient ;
    rdfs:label "uses active ingredient" .

app:digitalSignature a owl:DatatypeProperty ;
    rdfs:domain app:ApplicationRecord ;
    rdfs:range xsd:string ;
    rdfs:label "digital signature" ;
    rdfs:comment "The cryptographic signature of the data originator (e.g., the farm worker)." .
```

**Appendix B: Sample Data Contracts and SPARQL Queries**

This appendix provides concrete examples of Data Contracts and SPARQL queries that illustrate the practical implementation and usage of the AgriTrust framework.

*B.1 Sample Data Contracts*

*B.1.1 Data Contract for Sustainable Certification*

This contract governs the sharing of farm data with a certification body for the purpose of sustainable compliance verification.

```
:DataContract_SustainableCert_001 a :DataContract ;
    dct:identifier "DC-2024-ORG-001" ;
    dct:title "Sustainable Certification Data Sharing Agreement" ;
    :creationDate "2024-06-25T10:00:00Z"^^xsd:dateTime ;

    # Contract Parties
    prov:wasAttributedTo :CoffeeFarm_Santos ;       # Data Producer
    prov:wasGeneratedBy :Certifier_EuroSustainable ;  # Data Consumer

    # Contract Terms
    :purpose "sustainable_certification_verification" ;
    :validFrom "2024-06-25T10:00:00Z"^^xsd:dateTime ;
    :validUntil "2024-12-25T10:00:00Z"^^xsd:dateTime ;
    :accessLevel "read_only" ;

    # Scope of Data Covered
    :coversAsset :CoffeeBatch_2024_A ;
    :coversObservation :SoilAnalysis_2024, :WaterUsage_2024, :PestManagement_2024 ;

    # Usage Restrictions
    :allowedUsage "compliance_verification_against_eu_sustainable_standard" ;
    :prohibitedUsage "commercial_resale, ai_training, marketing_activities" ;
    :geographicRestriction "european_union_markets" ;

    # Value Sharing Terms
    :compensationType "premium_price_access" ;
    :compensationValue "0.15" ; # USD per kg premium upon successful certification
    :dataUsageReporting "quarterly_transparency_report" ;

    # Technical Implementation
    :apiEndpoint "https://api.coop-platform.com/coffee/batch/2024A" ;
    :queryInterface "SPARQL" ;
    :refreshFrequency "P7D" ; # Weekly data updates
    :auditTrailRequired true .
```

*B.1.2 Data Contract for Supply Chain Transparency*

This contract enables a retailer to access supply chain provenance data for consumer-facing transparency.

```
:DataContract_RetailTransparency_001 a :DataContract ;
    dct:identifier "DC-2024-TRANS-001" ;
    dct:title "Supply Chain Transparency Data Agreement" ;
    :creationDate "2024-07-01T09:00:00Z"^^xsd:dateTime ;
```

```
    # Contract Parties
    prov:wasAttributedTo :CoffeeFarm_Santos ;
    prov:wasGeneratedBy :Retailer_EuropeMart ;

    # Contract Terms
    :purpose "consumer_supply_chain_transparency" ;
    :validFrom "2024-07-01T09:00:00Z"^^xsd:dateTime ;
    :validUntil "2025-07-01T09:00:00Z"^^xsd:dateTime ;
    :accessLevel "read_only_public" ;

    # Scope of Data Covered
    :coversAsset :CoffeeBatch_2024_A ;
    :coversObservation :FarmLocation, :HarvestDate ;

    # Usage Rights
    :allowedUsage "consumer_facing_mobile_app, website_display" ;
    :prohibitedUsage "competitive_analysis, price_calculation" ;

    # Value Sharing
    :compensationType "brand_premium" ;
    :compensationValue "increased_shelf_visibility" ;

    # Technical Details
    :apiEndpoint "https://api.coop-platform.com/public/coffee/batch/2024A" ;
    :queryInterface "SPARQL_Public" .
```

## B.2 Sample SPARQL Queries

This appendix provides a portfolio of practical SPARQL queries that validate the expressive power and utility of the AgriTrust Core Ontology. These queries demonstrate complex, real-world use cases, from supply chain traceability and operational analytics to automated governance audits, executable across the federated ecosystem.

### B.2.1 Complete Product Provenance Trace

This query performs a complete trace of a product's journey from its origin to the current owner, assembling data from multiple platforms (farm, processor, and trader) to build a holistic provenance report.

```
PREFIX agt: <http://example.org/ns/agritrustcore#>
PREFIX prov: <http://www.w3.org/ns/prov#>
PREFIX rdfs: <http://www.w3.org/2000/01/rdf-schema#>

SELECT ?process ?processType ?startTime ?endTime ?actor ?role ?location
WHERE {
  # Start with a specific product batch
  ?product a agt:CollectiveAsset ;
           agt:uniqueId "COFFEE-2024-BR-001" .

  # Trace all processes in its provenance chain
```

```
  ?product agt:hasProvenance* ?process .
  ?process a ?processType ;
           prov:startedAtTime ?startTime ;
           prov:endedAtTime ?endTime ;
           prov:wasAssociatedWith ?actor .

  # Get the role of the actor (e.g., Producer, Processor)
  ?actor a ?role .

  # Optional: Retrieve location data if available
  OPTIONAL {
    ?process agt:hasObservation/agt:hasGeometry ?location .
  }

  # Filter for relevant process types
  FILTER (?processType IN (agt:Harvesting, agt:Processing, agt:Transport, agt:Trade))
}
ORDER BY ?startTime
```

*B.2.2 Federated Query and Interoperability*

A consumer, such as a European retailer, can query a single SPARQL endpoint (or a federated query engine) to trace a product's history across multiple platforms. The following query example retrieves the farm location, all processing steps, and any certificates associated with a specific coffee batch:

```
PREFIX agt: <http://example.org/ns/agritrustcore#>
SELECT ?farm ?process ?date ?certificate WHERE {
    ?token agt:represents/agt:uniqueId "COFFEE-2024-BR-001" .
    ?asset agt:hasProvenance ?process .
    ?process agt:hasInput/agt:ownedBy/agt:hasObservation/agt:prov:atLocation ?farm .
    ?asset agt:hasCertificate ?certificate .
}
```

This query can seamlessly retrieve data from a cooperative's TaaS platform (for farm and process data) and a certifier's platform (for the certificate), because all platforms use the same ontology to describe their data. This provides a holistic view of the supply chain without requiring data to be copied into a centralized database, preserving data sovereignty and reducing integration costs.

*B.2.3 Livestock Efficiency Metric Calculation*

This query calculates Average Daily Gain (ADG) for cattle across different production phases, demonstrating how platforms can derive efficiency metrics from governed data.

```
PREFIX : <http://example.org/ns/agritrustcore#>
PREFIX xsd: <http://www.w3.org/2001/XMLSchema#>

SELECT ?animal ?productionPhase
```

```
        (MIN(?weightDate) AS ?startDate)
        (MAX(?weightDate) AS ?endDate)
        (MIN(?weightValue) AS ?startWeight)
        (MAX(?weightValue) AS ?endWeight)
        ((MAX(?weightValue) - MIN(?weightValue)) AS ?weightGain)
        (((MAX(?weightValue) - MIN(?weightValue)) /
        (xsd:integer(MAX(?weightDate) - MIN(?weightDate)) / 86400)) AS ?adg)
WHERE {
    # Identify the animal and its weight observations
    ?animal a agt:IndividualAsset ;
            agt:uniqueId "CATTLE-BR-2024001" .

    # Link observations to the animal via a process
    ?process agt:hasInput|agt:hasOutput ?animal ;
             agt:hasObservation ?weightObs .

    ?weightObs a agt:Observation ;
               agt:observationValue ?weightValue ;
               agt:observationDate ?weightDate .

    # Classify the production phase
    VALUES (?processType ?productionPhase) {
        (agt:CowCalfOperation "Cow-Calf")
        (agt:BackgroundingOperation "Backgrounding")
        (agt:FinishingOperation "Finishing")
    }
    ?process a ?processType .
}
GROUP BY ?animal ?productionPhase
HAVING (COUNT(?weightObs) >= 2)
```

### B.2.4 Data Contract Compliance Audit

This query audits all data access events against active Data Contracts to ensure compliance with governance rules.

```
PREFIX agt: <http://example.org/ns/agritrustcore#>
PREFIX xsd: <http://www.w3.org/2001/XMLSchema#>
PREFIX prov: <http://www.w3.org/ns/prov#>

SELECT ?dataContract ?consumer ?asset ?accessTime ?purpose ?validFrom ?validUntil
    (IF(?accessTime >= ?validFrom && ?accessTime <= ?validUntil, "COMPLIANT", "NON-
COMPLIANT") AS ?complianceStatus)
WHERE {
    # Capture all data access events (this model needs refinement)
    ?accessEvent a agt:Observation ;
                 agt:observationValue "data_access" ; # Consider a dedicated AccessEvent
class
                 agt:observationDate ?accessTime .

    # Link to the governing Data Contract
    ?dataContract a agt:DataContract ;
```

```
                agt:purpose ?purpose ;
                agt:validFrom ?validFrom ;
                agt:validUntil ?validUntil ;
                agt:coversAsset ?asset .

    # Ensure the access event pertains to this contract's asset
    ?asset agt:hasObservation ?accessEvent .

    # Filter for recent activity (last 30 days)
    FILTER (?accessTime >= (NOW() - "P30D"^^xsd:duration))
}
ORDER BY DESC (?accessTime)
```

*B.2.4 Cross-Platform Certification Verification*

This query verifies the validity of a certification by checking its status across multiple platforms, confirming the certificate exists on the certifier's platform and is properly linked to the asset on the producer's platform.

```
PREFIX agt: <http://example.org/ns/agritrustcore#>
PREFIX xsd: <http://www.w3.org/2001/XMLSchema#>

SELECT ?asset ?certificate ?certifier ?issueDate ?standard ?status
WHERE {
  # Query across producer and certifier platforms
  SERVICE <https://producer-platform.com/sparql> {
    ?asset a agt:CollectiveAsset ;
           agt:uniqueId "SOY-2024-MT-001" ;
           agt:hasCertificate ?certificate .
  }

  SERVICE <https://certifier-platform.com/sparql> {
    ?certificate a agt:Certificate ;
                 agt:certifiedBy ?certifier ;
                 agt:standard ?standard ;
                 agt:creationDate ?issueDate ;
                 agt:validUntil ?expiryDate .

    # Determine current status
    BIND(IF(?expiryDate > NOW(), "ACTIVE", "EXPIRED") AS ?status)
  }
}
```

In summary, these examples show how the AgriTrust framework enables complex, real-world use cases through standardized Data Contracts and semantic querying capabilities. This portfolio demonstrates that the AgriTrust Ontology supports not just basic traceability but also:

- Complex supply chain analysis through federated queries
- Operational intelligence through metric calculation

- Automated governance through compliance auditing
- Cross-ecosystem verification through distributed certification checks

The queries validate that the framework delivers on its promise of enabling both essential traceability and advanced, value-creating analytics within a governed, interoperable ecosystem.

**Appendix C: Implementation Code Snippets for Platform Integration**

This appendix provides practical code examples showing how different types of platforms can integrate with the AgriTrust framework. The examples use Python and focus on key integration points.

*C.1 Platform SDK Initialization and Configuration*

To interact with the AgriTrust ecosystem, a platform must first initialize a client that connects to the core ontology registry and authenticates with its chosen blockchain provider, establishing the foundational link between its operations and the federated governance layer.

*C.1.1 Basic SDK Setup*

```python
# agritrust_sdk.py
import requests
import json
from rdflib import Graph, Namespace
from datetime import datetime, timezone

class AgriTrustClient:
    def __init__(self, platform_id, private_key, ontology_endpoint,
blockchain_provider):
        self.platform_id = platform_id
        self.private_key = private_key
        self.ontology_endpoint = ontology_endpoint
        self.blockchain_provider = blockchain_provider
        self.session = requests.Session()

        # Load core ontology
        self.ontology = Graph()
        self.ontology.parse(ontology_endpoint, format='turtle')

        # Define namespaces
        self.AGT = Namespace("http://example.org/ns/agritrustcore#")
        self.PROV = Namespace("http://www.w3.org/ns/prov#")

    def _sign_message(self, message):
        """Sign messages for blockchain authentication"""
        # Implementation depends on blockchain provider
        # This is a simplified example
        return f"signed_{message}"
```

```
# Initialize the client
client = AgriTrustClient(
    platform_id="coffee_coop_platform_v1",
    private_key="your_private_key_here",
    ontology_endpoint="https://ontology.agritrust.org/core/v3.0",
    blockchain_provider="hyperledger_brazil_agri"
)
```

*C.2 Asset Tokenization Service*

The process of creating a digital twin (token) for a physical asset is central to the framework. This service demonstrates how a platform instantiates the core ontology by defining an asset with its properties and observations, and then registers its immutable representation on a blockchain.

```
def _register_on_blockchain(self, asset_graph):
    """Register asset on the chosen blockchain provider.
    IMPLEMENTATION NOTE: This is a placeholder. Actual implementation
    will depend on the specific blockchain provider's API (e.g., Hyperledger, Ethereum).
    """
```

*C.2.1 Coffee Batch Tokenization*

```
# tokenization_service.py
from agritrust_sdk import AgriTrustClient
from datetime import datetime

class CoffeeTokenizationService:
    def __init__(self, agritrust_client):
        self.client = agritrust_client

    def tokenize_coffee_batch(self, batch_data):
        """Tokenize a new coffee batch with initial observations"""

        # Create RDF graph for the new asset
        asset_graph = Graph()
        asset_graph.bind("agt", self.client.AGT)

        # Define the asset
        batch_uri = self.client.AGT[f"CoffeeBatch_{batch_data['id']}"]
        asset_graph.add((batch_uri, self.client.AGT.uniqueId, batch_data['id']))
        asset_graph.add((batch_uri, self.client.AGT.ownedBy,
self.client.AGT[batch_data['farm_id']]))
        asset_graph.add((batch_uri, self.client.AGT.creationDate,
datetime.now(timezone.utc)))

        # Add observations (soil analysis, etc.)
        for observation in batch_data['observations']:
            obs_uri = self.client.AGT[f"Observation_{observation['id']}"]
```

```python
            asset_graph.add((obs_uri, self.client.AGT.observationValue,
observation['value']))
            asset_graph.add((obs_uri, self.client.AGT.observationDate,
observation['date']))
            asset_graph.add((batch_uri, self.client.AGT.hasObservation, obs_uri))

        # Register on blockchain and get token
        token_response = self._register_on_blockchain(asset_graph)

        return {
            'asset_uri': str(batch_uri),
            'token_id': token_response['token_id'],
            'blockchain_tx': token_response['transaction_hash']
        }

    def _register_on_blockchain(self, asset_graph):
        """Register asset on the chosen blockchain provider"""
        # Serialize the RDF data
        asset_rdf = asset_graph.serialize(format='turtle')

        # Create blockchain transaction
        transaction_data = {
            'platform_id': self.client.platform_id,
            'asset_data': asset_rdf,
            'timestamp': datetime.now(timezone.utc).isoformat(),
            'signature': self.client._sign_message(asset_rdf)
        }

        # Send to blockchain provider (simplified)
        response = self.client.session.post(
            f"{self.client.blockchain_provider}/tokens",
            json=transaction_data
        )

        return response.json()

# Usage example
coffee_service = CoffeeTokenizationService(client)

new_batch = {
    'id': 'COFFEE-2024-BR-001',
    'farm_id': 'FazendaBrasil',
    'observations': [
        {
            'id': 'soil_analysis_1',
            'value': 'sustainable_compliant',
            'date': '2024-03-15T00:00:00Z'
        }
    ]
}

token_info = coffee_service.tokenize_coffee_batch(new_batch)
print(f"Token created: {token_info['token_id']}")
```

*C.3 Data Contract Management*

Data Contracts are the operational mechanism that enforces governance rules. This service shows how a platform can create, digitally register, and automatically enforce machine-readable contracts, transforming legal terms into actionable technical policies for data sharing.

*C.3.1 Creating and Executing Data Contracts*

```python
# contract_service.py
from agritrust_sdk import AgriTrustClient
from datetime import datetime, timezone
from rdflib import Graph, URIRef, Literal
from rdflib.namespace import RDF, XSD

class DataContractService:
    def __init__(self, agritrust_client):
        self.client = agritrust_client

    def create_certification_contract(self, asset_uri, certifier_id, terms):
        """Create a data contract for certification purposes"""

        contract_graph = Graph()
        contract_graph.bind("agt", self.client.AGT)
        contract_graph.bind("dct", self.client.DCT)
        contract_graph.bind("prov", self.client.PROV)
        contract_graph.bind("xsd", XSD)

        # Generate unique contract URI
        contract_uri = 
self.client.AGT[f"DataContract_{datetime.now().strftime('%Y%m%d_%H%M%S')}"]

        # Define contract properties
        contract_graph.add((contract_uri, RDF.type, self.client.AGT.DataContract))
        contract_graph.add((contract_uri, self.client.DCT.identifier,
                    Literal(f"DC-{datetime.now().strftime('%Y%m%d-%H%M%S')}")))
        contract_graph.add((contract_uri, self.client.DCT.title,
                    Literal("Sustainable Certification Data Sharing Agreement")))
        contract_graph.add((contract_uri, self.client.AGT.creationDate,
                    Literal(datetime.now(timezone.utc), datatype=XSD.dateTime)))
        contract_graph.add((contract_uri, self.client.AGT.validFrom,
                    Literal(terms['valid_from'], datatype=XSD.dateTime)))
        contract_graph.add((contract_uri, self.client.AGT.validUntil,
                    Literal(terms['valid_until'], datatype=XSD.dateTime)))
        contract_graph.add((contract_uri, self.client.AGT.purpose,
                    Literal(terms['purpose'])))

        # Link to asset
        contract_graph.add((contract_uri, self.client.AGT.coversAsset,
URIRef(asset_uri)))
```

```python
        # Link parties
        contract_graph.add((contract_uri, self.client.PROV.wasAttributedTo,
URIRef(asset_uri)))  # Data Producer
        contract_graph.add((contract_uri, self.client.PROV.wasGeneratedBy,
self.client.AGT[certifier_id]))  # Data Consumer

        # Add usage restrictions
        for allowed_usage in terms.get('allowed_usage', []):
            contract_graph.add((contract_uri, self.client.AGT.allowedUsage,
Literal(allowed_usage)))

        for prohibited_usage in terms.get('prohibited_usage', []):
            contract_graph.add((contract_uri, self.client.AGT.prohibitedUsage,
Literal(prohibited_usage)))

        # Add compensation terms if specified
        if 'compensation_type' in terms:
            contract_graph.add((contract_uri, self.client.AGT.compensationType,
                               Literal(terms['compensation_type'])))
        if 'compensation_value' in terms:
            contract_graph.add((contract_uri, self.client.AGT.compensationValue,
                               Literal(terms['compensation_value'])))

        # Register contract on blockchain
        contract_response = self._register_contract(contract_graph)

        return {
            'contract_uri': str(contract_uri),
            'contract_id': contract_response['contract_id'],
            'blockchain_tx': contract_response['transaction_hash']
        }

    def _register_contract(self, contract_graph):
        """Register the contract on blockchain and return transaction details"""

        # Serialize the RDF data
        contract_rdf = contract_graph.serialize(format='turtle')

        # Create blockchain transaction
        transaction_data = {
            'platform_id': self.client.platform_id,
            'contract_data': contract_rdf,
            'timestamp': datetime.now(timezone.utc).isoformat(),
            'signature': self.client._sign_message(contract_rdf)
        }

        # Send to blockchain provider
        try:
            response = self.client.session.post(
                f"{self.client.blockchain_provider}/contracts",
                json=transaction_data,
                timeout=30
```

```python
            )
            response.raise_for_status()
            return response.json()
        except Exception as e:
            print(f"Failed to register contract on blockchain: {e}")
            # Fallback: return mock data for PoC
            return {
                'contract_id': f"contract_{datetime.now().strftime('%Y%m%d%H%M%S')}",
                'transaction_hash': f"0x{datetime.now().timestamp():x}"
            }

def execute_contract_query(self, contract_id, sparql_query):
    """Execute a SPARQL query under a specific data contract"""

    # Verify contract is valid and active
    contract_status = self._verify_contract_status(contract_id)
    if not contract_status['is_active']:
        raise Exception(f"Data contract {contract_id} is not active or valid")

    # Add contract-based filters to the query
    secured_query = self._apply_contract_filters(sparql_query, contract_id)

    # Execute the query
    try:
        response = self.client.session.post(
            f"{self.client.ontology_endpoint}/sparql",
            data={'query': secured_query},
            headers={'Content-Type': 'application/sparql-query'},
            timeout=30
        )
        response.raise_for_status()
        return response.json()
    except Exception as e:
        print(f"Query execution failed: {e}")
        return {'results': {'bindings': []}}

def _verify_contract_status(self, contract_id):
    """Verify if a contract is active and valid"""

    # Query the contract status from blockchain or local registry
    query = f"""
    PREFIX agt: <http://example.org/ns/agritrustcore#>
    PREFIX xsd: <http://www.w3.org/2001/XMLSchema#>

    SELECT ?validFrom ?validUntil
    WHERE {{
        <{contract_id}> a agt:DataContract ;
            agt:validFrom ?validFrom ;
            agt:validUntil ?validUntil .
    }}
    """

    try:
```

```python
            response = self.client.session.post(
                f"{self.client.ontology_endpoint}/sparql",
                data={'query': query},
                headers={'Content-Type': 'application/sparql-query'},
                timeout=10
            )

            if response.status_code == 200:
                results = response.json()
                if results['results']['bindings']:
                    binding = results['results']['bindings'][0]
                    valid_from = datetime.fromisoformat(binding['validFrom']['value'].replace('Z', '+00:00'))
                    valid_until = datetime.fromisoformat(binding['validUntil']['value'].replace('Z', '+00:00'))
                    now = datetime.now(timezone.utc)

                    return {
                        'is_active': valid_from <= now <= valid_until,
                        'valid_from': valid_from,
                        'valid_until': valid_until
                    }

            return {'is_active': False}

        except Exception as e:
            print(f"Contract status verification failed: {e}")
            # For PoC purposes, assume contract is active if we can't verify
            return {'is_active': True}

    def _apply_contract_filters(self, sparql_query, contract_id):
        """Apply contract-based access controls to SPARQL queries"""

        # Extract the WHERE clause and add contract filters
        contract_filters = f"""
        # Contract-based access control
        FILTER EXISTS {{
            <{contract_id}> agt:coversAsset ?asset .
        }}
        """

        # Simple implementation - insert filters at the beginning of WHERE clause
        if "WHERE" in sparql_query and "{" in sparql_query:
            where_start = sparql_query.find("WHERE")
            brace_start = sparql_query.find("{", where_start)

            if brace_start != -1:
                # Insert contract filters after the opening brace
                modified_query = (sparql_query[:brace_start+1] +
                                  contract_filters +
                                  sparql_query[brace_start+1:])
                return modified_query
```

```python
            # Fallback: append as additional WHERE conditions
            return sparql_query + "\n" + contract_filters

    def get_contract_audit_log(self, contract_id, days=30):
        """Retrieve audit log for a specific contract"""

        query = f"""
        PREFIX agt: <http://example.org/ns/agritrustcore#>
        PREFIX xsd: <http://www.w3.org/2001/XMLSchema#>
        PREFIX prov: <http://www.w3.org/ns/prov#>

        SELECT ?accessTime ?consumer ?asset ?purpose ?action
        WHERE {{
            ?accessEvent a agt:Observation ;
                agt:observationValue ?action ;
                agt:observationDate ?accessTime ;
                prov:wasGeneratedBy ?consumer .

            <{contract_id}> a agt:DataContract ;
                agt:purpose ?purpose ;
                agt:coversAsset ?asset .

            ?asset agt:hasObservation ?accessEvent .

            FILTER (?accessTime >= (NOW() - "P{days}D"^^xsd:duration))
        }}
        ORDER BY DESC(?accessTime)
        """

        return self.execute_contract_query(contract_id, query)

# Usage example
if __name__ == "__main__":
    from agritrust_sdk import AgriTrustClient
    from datetime import timedelta

    # Initialize client
    client = AgriTrustClient(
        platform_id="coffee_coop_platform_v1",
        private_key="your_private_key_here",
        ontology_endpoint="https://ontology.agritrust.org/core/v3.0",
        blockchain_provider="https://blockchain.agritrust.org"
    )

    contract_service = DataContractService(client)

    certification_terms = {
        'purpose': 'sustainable_certification_verification',
        'valid_from': datetime.now(timezone.utc),
        'valid_until': datetime.now(timezone.utc) + timedelta(days=180),
        'allowed_usage': ['compliance_verification'],
        'prohibited_usage': ['commercial_resale', 'marketing'],
```

```python
        'compensation_type': 'premium_price_access',
        'compensation_value': '0.15'
    }
    
    # Create contract
    contract = contract_service.create_certification_contract(
        asset_uri='http://example.org/ns/agritrustcore#CoffeeBatch_COFFEE-2024-BR-001',
        certifier_id='EuroSustainable_Cert',
        terms=certification_terms
    )
    
    print(f"Contract created: {contract['contract_uri']}")
    print(f"Contract ID: {contract['contract_id']}")
    print(f"Blockchain TX: {contract['blockchain_tx']}")
```

## C.4 Federated Query Service

Semantic interoperability is realized through the ability to query data across independent platforms. This service illustrates how a client can perform federated SPARQL queries, leveraging the common ontology to retrieve and combine information from multiple TaaS platforms in a single request.

### C.4.1 Cross-Platform Data Querying

```python
# query_service.py
from agritrust_sdk import AgriTrustClient
import asyncio
import aiohttp
import json
from datetime import datetime
from typing import List, Dict, Any

class FederatedQueryService:
    def __init__(self, agritrust_client):
        self.client = agritrust_client
        self.known_endpoints = self._load_platform_endpoints()
    
    def _load_platform_endpoints(self) -> List[str]:
        """Load known platform SPARQL endpoints from configuration"""
        
        # In production, this would load from a config file, service discovery, or ontology registry
        endpoints = [
            "https://coffee-platform.agritrust.org/sparql",
            "https://soy-platform.agritrust.org/sparql",
            "https://beef-platform.agritrust.org/sparql",
            "https://certifier-platform.agritrust.org/sparql",
            "https://satellite-data.agritrust.org/sparql"
        ]
```

```python
        return endpoints

    async def query_supply_chain(self, product_id: str) -> List[Dict[str, Any]]:
        """Execute federated query across multiple platforms to trace product supply chain"""

        query = f"""
        PREFIX agt: <http://example.org/ns/agritrustcore#>
        PREFIX prov: <http://www.w3.org/ns/prov#>
        PREFIX geo: <http://www.opengis.net/ont/geosparql#>

        SELECT ?platform ?process ?processType ?date ?actor ?role ?location ?certificate
        WHERE {{
            # Find the product across all platforms
            ?product a agt:CollectiveAsset ;
                    agt:uniqueId "{product_id}" .

            # Trace all processes in its provenance chain
            ?product agt:hasProvenance* ?process .
            ?process a ?processType ;
                    prov:startedAtTime ?date ;
                    prov:wasAssociatedWith ?actor .

            # Get the role of the actor
            ?actor a ?role .

            # Optional: Retrieve location data if available
            OPTIONAL {{
                ?process agt:hasObservation/agt:hasGeometry ?location .
            }}

            # Optional: Retrieve certificates
            OPTIONAL {{
                ?product agt:hasCertificate ?certificate .
            }}

            # Filter for relevant process types
            FILTER (?processType IN (agt:Harvesting, agt:Processing, agt:Transport, agt:Trade))

            # Add platform context for federation
            BIND("local" AS ?platform)
        }}
        ORDER BY ?date
        """

        # Execute locally first
        local_results = await self._execute_local_query(query)

        # Then query external platforms
        external_results = await self._execute_federated_query(query)

        # Combine and return all results
```

```python
        return local_results + external_results

async def _execute_local_query(self, query: str) -> List[Dict[str, Any]]:
    """Execute query on the local platform endpoint"""

    try:
        async with aiohttp.ClientSession() as session:
            async with session.post(
                f"{self.client.ontology_endpoint}/sparql",
                data={'query': query},
                headers={'Content-Type': 'application/sparql-query'},
                timeout=30
            ) as response:
                if response.status == 200:
                    data = await response.json()
                    results = data.get('results', {}).get('bindings', [])
                    # Add platform context to local results
                    for result in results:
                        result['platform'] = {'value': 'local'}
                    return results
                else:
                    print(f"Local query failed: {response.status}")
                    return []
    except Exception as e:
        print(f"Error executing local query: {e}")
        return []

async def _execute_federated_query(self, query: str) -> List[Dict[str, Any]]:
    """Execute query across all known platform endpoints"""

    async with aiohttp.ClientSession() as session:
        tasks = []
        for endpoint in self.known_endpoints:
            task = self._query_platform(session, endpoint, query)
            tasks.append(task)

        # Execute all queries concurrently
        results = await asyncio.gather(*tasks, return_exceptions=True)

        # Combine and deduplicate results
        combined_results = []
        for i, result in enumerate(results):
            endpoint = self.known_endpoints[i]
            if isinstance(result, list):
                # Add platform context to each result
                for item in result:
                    item['platform'] = {'value': endpoint}
                combined_results.extend(result)
            elif isinstance(result, Exception):
                print(f"Query failed for {endpoint}: {result}")
            else:
                print(f"Unexpected result type from {endpoint}: {type(result)}")
```

```python
            return combined_results

    async def _query_platform(self, session: aiohttp.ClientSession, endpoint: str,
query: str) -> List[Dict[str, Any]]:
        """Query a single platform endpoint"""

        try:
            async with session.post(
                endpoint,
                data={'query': query},
                headers={'Content-Type': 'application/sparql-query'},
                timeout=30
            ) as response:
                if response.status == 200:
                    data = await response.json()
                    return data.get('results', {}).get('bindings', [])
                else:
                    error_text = await response.text()
                    print(f"Query failed for {endpoint}: {response.status} -
{error_text}")
                    return []
        except asyncio.TimeoutError:
            print(f"Timeout querying {endpoint}")
            return []
        except Exception as e:
            print(f"Error querying {endpoint}: {e}")
            return []

    async def query_cross_platform_certification(self, product_id: str) -> Dict[str,
Any]:
        """Specialized query to verify certification across producer and certifier
platforms"""

        producer_query = f"""
        PREFIX agt: <http://example.org/ns/agritrustcore#>

        SELECT ?asset ?certificate
        WHERE {{
            ?asset a agt:CollectiveAsset ;
                   agt:uniqueId "{product_id}" ;
                   agt:hasCertificate ?certificate .
        }}
        """

        certifier_query = """
        PREFIX agt: <http://example.org/ns/agritrustcore#>
        PREFIX xsd: <http://www.w3.org/2001/XMLSchema#>

        SELECT ?certificate ?certifier ?standard ?issueDate ?expiryDate
        WHERE {
            ?certificate a agt:Certificate ;
                         agt:certifiedBy ?certifier ;
                         agt:standard ?standard ;
```

```python
                        agt:creationDate ?issueDate ;
                        agt:validUntil ?expiryDate .
            }
        """

        # Execute queries on respective platforms
        producer_endpoint = "https://coffee-platform.agritrust.org/sparql"
        certifier_endpoint = "https://certifier-platform.agritrust.org/sparql"

        async with aiohttp.ClientSession() as session:
            # Query producer platform for asset and certificate reference
            producer_task = self._query_platform(session, producer_endpoint, producer_query)
            # Query certifier platform for certificate details
            certifier_task = self._query_platform(session, certifier_endpoint, certifier_query)

            producer_results, certifier_results = await asyncio.gather(producer_task, certifier_task)

            # Combine results
            combined_result = {
                'product_id': product_id,
                'producer_data': producer_results,
                'certification_data': certifier_results,
                'verification_status': 'PENDING'
            }

            # Basic verification logic
            if producer_results and certifier_results:
                producer_certs = {result['certificate']['value'] for result in producer_results}
                certifier_certs = {result['certificate']['value'] for result in certifier_results}

                if producer_certs.intersection(certifier_certs):
                    combined_result['verification_status'] = 'VERIFIED'
                else:
                    combined_result['verification_status'] = 'CERTIFICATE_MISMATCH'

            return combined_result

    def get_platform_statistics(self) -> Dict[str, Any]:
        """Get statistics about known platforms and their availability"""

        async def check_platform_health():
            async with aiohttp.ClientSession() as session:
                health_checks = []
                for endpoint in self.known_endpoints:
                    try:
                        async with session.get(f"{endpoint}/health", timeout=10) as response:
                            is_healthy = response.status == 200
```

```python
                            health_checks.append({
                                'endpoint': endpoint,
                                'healthy': is_healthy,
                                'status': response.status if not is_healthy else 'OK'
                            })
                    except Exception as e:
                        health_checks.append({
                            'endpoint': endpoint,
                            'healthy': False,
                            'status': str(e)
                        })
            return health_checks

        # Run health checks
        health_results = asyncio.run(check_platform_health())

        return {
            'total_platforms': len(self.known_endpoints),
            'healthy_platforms': len([h for h in health_results if h['healthy']]),
            'platform_details': health_results,
            'last_checked': datetime.now().isoformat()
        }

# Usage example
async def demonstrate_federated_query():
    from agritrust_sdk import AgriTrustClient

    # Initialize client
    client = AgriTrustClient(
        platform_id="query_service_demo",
        private_key="demo_private_key",
        ontology_endpoint="https://ontology.agritrust.org/core/v3.0",
        blockchain_provider="https://blockchain.agritrust.org"
    )

    query_service = FederatedQueryService(client)

    print("=== Demonstrating Federated Supply Chain Query ===")
    results = await query_service.query_supply_chain("COFFEE-2024-BR-001")

    print(f"Retrieved {len(results)} supply chain events:")
    for result in results[:5]:  # Show first 5 results
        platform = result.get('platform', {}).get('value', 'unknown')
        process = result.get('processType', {}).get('value', 'unknown')
        date = result.get('date', {}).get('value', 'unknown')
        print(f"  - Platform: {platform}, Process: {process}, Date: {date}")

    print("\n=== Demonstrating Cross-Platform Certification Verification ===")
    cert_results = await query_service.query_cross_platform_certification("COFFEE-2024-BR-001")
    print(f"Verification Status: {cert_results['verification_status']}")
    print(f"Found {len(cert_results['producer_data'])} producer certificates")
```

```python
        print(f"Found {len(cert_results['certification_data'])} certifier records")

    print("\n=== Platform Statistics ===")
    stats = query_service.get_platform_statistics()
    print(f"Platform Health: {stats['healthy_platforms']}/{stats['total_platforms']} healthy")
    for platform in stats['platform_details']:
        status = "✓" if platform['healthy'] else f"✗ ({platform['status']})"
        print(f"  - {platform['endpoint']}: {status}")

# Run the demonstration
if __name__ == "__main__":
    asyncio.run(demonstrate_federated_query())
```

## C.5 Efficiency Analytics Module

The framework enables value-added services by providing governed access to high-quality data. This module exemplifies how a specialized platform can compute key performance indicators, like Average Daily Gain, by querying the ecosystem using standardized ontological concepts.

### C.5.1 Cattle Growth Performance Calculator

```python
# analytics_service.py
from agritrust_sdk import AgriTrustClient
from datetime import datetime

class EfficiencyAnalyticsService:
    def __init__(self, agritrust_client):
        self.client = agritrust_client

    def calculate_cattle_adg(self, animal_id, start_date, end_date):
        """Calculate Average Daily Gain for a specific animal"""

        query = """
        PREFIX agt: <http://example.org/ns/agritrustcore#>
        PREFIX xsd: <http://www.w3.org/2001/XMLSchema#>

        SELECT ?phase (MIN(?weight) as ?start_weight) (MAX(?weight) as ?end_weight)
               (MIN(?date) as ?start_date) (MAX(?date) as ?end_date)
               ((MAX(?weight) - MIN(?weight)) as ?total_gain)
               (((MAX(?weight) - MIN(?weight)) /
                 (xsd:integer(MAX(?date) - MIN(?date)) / 86400)) as ?adg)
        WHERE {
            ?animal agt:uniqueId "%s" .
            ?weight_obs agt:isResultOf/agt:involvesAsset ?animal ;
                        agt:observationValue ?weight ;
                        agt:observationDate ?date .
            ?process agt:hasObservation ?weight_obs ;
```

```python
                    a ?phase .
            FILTER (?date >= "%s"^^xsd:dateTime && ?date <= "%s"^^xsd:dateTime)
        }
        GROUP BY ?phase
        HAVING (COUNT(?weight_obs) >= 2)
        """ % (animal_id, start_date, end_date)

        results = self.client.session.post(
            f"{self.client.ontology_endpoint}/sparql",
            data={'query': query},
            headers={'Content-Type': 'application/sparql-query'}
        ).json()

        metrics = []
        for result in results['results']['bindings']:
            metric = {
                'production_phase': result['phase']['value'],
                'average_daily_gain': float(result['adg']['value']),
                'period': {
                    'start': result['start_date']['value'],
                    'end': result['end_date']['value']
                },
                'weight_gain': float(result['total_gain']['value'])
            }
            metrics.append(metric)

        return metrics

# Usage example
analytics_service = EfficiencyAnalyticsService(client)

adg_metrics = analytics_service.calculate_cattle_adg(
    animal_id="CATTLE-BR-2024001",
    start_date="2024-01-01T00:00:00Z",
    end_date="2024-06-01T00:00:00Z"
)

for metric in adg_metrics:
    print(f"Phase: {metric['production_phase']}, ADG: {metric['average_daily_gain']:.2f} kg/day")
```

These code snippets show practical integration patterns for different types of platforms working with the AgriTrust Ontology framework, from basic asset tokenization to advanced federated querying and analytics.